\documentclass{aa}

\newif\ifpdf
\ifx \pdfoutput \undefined
\pdffalse                          
\else
\pdfoutput=1
\pdftrue
\fi

\ifpdf
\usepackage[pdftex]{graphicx}
\else
\usepackage{graphicx}
\fi

\begin{document}

\def \etal{{\it et\,al.\/}}
\def \privcom{{\it priv.\,comm.\/}}
\def \inprep{{\it in preparation\/}}
\def \eg{{\it e.g.\/}}
\def \cf{{\it cf.\/}}
\def \ie{{\it i.e.\/}}
\def \ibidem{\quad{\it ibidem\/} }
\def \dito{\vrule width 0.8cm height 2.5pt depth -2.2pt $\,$}
\def\hlinet{\hline\rule{0cm}{3ex}}
\def\hlineb{\hline\rule[-1ex]{0cm}{1ex}}
\def\hlinetb{\hline\rule[-1ex]{0cm}{4ex}}
\def \Kp{K$^\prime$}
\def \FP{Fabry-Perot}

\title{
The expected abundance of Lyman-$\alpha$ emitting primeval galaxies.\\ 
I. General model predictions. }

\author{
        E.\ Thommes \inst{1,2}
   \and K.\ Meisenheimer\inst{1}
}

\authorrunning{Thommes \& Meisenheimer}
\titlerunning{Primeval Lyman-$\alpha$ galaxies}

\offprints{Eduard Thommes (E.Thommes@thphys.uni-heidelberg.de)}

\institute{ Max--Planck--Institut f\"ur Astronomie (MPIA),
K\"onigstuhl \/17, D-69117 Heidelberg, Germany 
\and Institut f\"ur
Theoretische Physik, Universit\"at Heidelberg, Philosophenweg 16,\/D-69120 Heidelberg,
Germany }

\date{Received xxx ; accepted xxx }

\markboth{Primeval Ly-$\alpha$ galaxies}{Primeval Ly-$\alpha$ galaxies}

\abstract{
We present model calculations for the expected surface density of
Ly-$\alpha$ emitting primeval galaxies (PGs) at high redshifts.  We
assume that elliptical galaxies and bulges of spiral galaxies (=
spheroids) formed early in the universe and that the Ly-$\alpha$
emitting PGs are these spheroids during their first burst of star
formation at high redshift. One of the main assumptions of the models
is that the Ly-$\alpha$ bright phase of this first starburst in the
spheroids is confined to a short period after its onset due to rapid
formation of dust.  The models do not only explain the failure of
early surveys for Ly-$\alpha $ emitting PGs but are also consistent
with the limits of new surveys (e.g. the Calar Alto Deep Imaging
Survey - CADIS).  At faint detection limits $S_{\mbox{lim}} \le
10^{-20}$W/m$^2$ the surface density of Ly-$\alpha$ emitters is
expected to vary only weakly in the redshift range between $z=3$ and
$z=6$ with values $> 10^3 / \mbox{deg}^2 / \Delta z = 0.1 $ reaching
its maximum at $z_0 \approx 4$.  At shallower detection limits,
$S_{\mbox{lim}} \ge 3\times 10^{-20}$W/m$^2$ the surface density of
high-z Ly-$\alpha$ emitters is expected to be a steep function of
redshift and detection limit. This explains the low success in finding
bright Ly-$\alpha$ galaxies at $z > 5$.  We demonstrate how the
observed  surface
densities of Ly-$\alpha $ emitting PGs derived from recent surveys constrain
the parameters of our models. Finally, we discuss the possibility that
two Ly-$\alpha$ bright phases occur in the formation process of
galaxies: An initial -- primeval -- phase in which dust is virtually
non-existant, and a later secondary phase in which strong galactic
winds as observed in some Lyman break galaxies 
facilitate the escape of Ly-$\alpha $ photons after dust has
already been formed.

}

\maketitle

\section{Introduction}

The detection of the ancestors of large present day galaxies (like our
Milky Way) during their first phase of violent star formation is
currently one of the
great challenges for observational cosmology. Commonly, these
objects are referred to as primeval galaxies (PGs). Finding them in
substantial numbers and over a sufficiently broad range of luminosities
would provide us with direct insight into the epoch of galaxy formation in
the young universe. Since current wisdom places the PG phase of our Galaxy
between redshifts $z=5$ and $z=10$ one should ultimately aim
to establish the luminosity function of primeval galaxies and their
evolution at several redshifts within this range.

It was not long ago that such observational program would have
seemed audacious.  
But with half a dozen telescopes of the 8-10m class
in operation, the detection of star formation rates of ten
M$_{\sun}$/year  appears feasible even at $z > 5$. Indeed, the
last few years have witnessed an enormous progress in
identifying young galaxies at very high redshifts (see Dey \etal, \cite{Dey98},
Weymann \etal\, \cite{Weymann98}, Hu \etal\, \cite{Hu98}, \cite{Hu99}, 
\cite{Hu02}, \cite{Hu03}, Rhoads \etal\, \cite{Rhoads03}).

Various techniques have been used to search for galaxies at high redshifts.
Currently the most successful method, introduced by
 Steidel \etal\ (\cite{Steidel92}, \cite{Steidel93}, \cite{Steidel96a}, \cite{Steidel96b}, \cite{Steidel98a}, 
\cite{Steidel98b}) 
uses the Lyman break and the flat
spectral energy distribution blue-wards of the Lyman break as a signature
of very distant, young star-forming galaxies.
Hundreds of galaxies found in this way have been
spectroscopically confirmed to be young star forming galaxies at redshifts
$z \simeq 3$ (Steidel \etal\, \cite{Steidel96a}). 
In addition several dozens of them could be detected at $z \simeq 4$
(Steidel \etal\ \cite{Steidel99}). The number density and clustering properties of these Lyman break galaxies
are consistent with them being the central galaxies of the most
massive dark matter halos present at $z\sim 3$ (Mo \etal\ \cite{Mo99}).
Although the Lyman break galaxies seem to be relatively young star forming galaxies,
they cannot represent the population of primeval galaxies in the sense
defined above: The strong metal absorption lines in 
their spectra  indicate that their current star formation
period must have been preceded by an earlier star burst. The massive stars
formed have already substantially enriched the interstellar medium
in these systems. Moreover, both the UV continuum slope and the Balmer
lines (Pettini \etal \cite{Pettini98}) suggest that the Lyman-break
galaxies already contain significant amounts of dust. Therefore, it is now
widely accepted that the star forming rates inferred from the UV continua
of Lyman-break galaxies have to be corrected upwards by a factor of 2
to 5. 
The presence of dust also
explains why only about 1/3 of the Lyman-break galaxies exhibit a strong
Ly-$\alpha$ line: In typical star forming regions every Ly-$\alpha$ photon emitted by
the hot OB stars will undergo dozens of multiple resonant scatterings
before leaving the region towards the observer. Thus even small amounts of
dust could quench the escaping Ly-$\alpha$ emission considerably. 

In contrast to the Lyman-break galaxies, all galaxies known at $z>5$
exhibit a very strong Ly-$\alpha$ emission line (rest frame equivalent
widths $> 5$\, nm). This is exactly the spectral signature we expect
for the first few hundred million years after the onset of a violent burst of
massive star formation and before the newly produced metals could be
recycled into the cold phase of the inter-stellar medium. Note that at
$z = 3$ there still exists a large population of Ly-$\alpha$ bright
galaxies (Shapley \etal\ \cite{Shapley03}, Kudritzki \etal\  \cite{Kudritzki00}) 
but their low star forming
rates indicate that they represent a population of smaller galaxies in
which the "trigger density" has been reached later than in the
Lyman-break systems. The weak continua of Ly-$\alpha $ bright
galaxies make them hard to find by color-break techniques, but the strong
emission lines should easily be detected in narrow-band searches for
emission line objects.

Although early attempts to detect Ly-$\alpha$ bright galaxies at $z
\ga 3$ failed (Pritchet \cite{Pritchet94}), we have witnessed a breakthrough in
finding these objects at redshifts between 3 and 5 as 
sufficiently deep detection limits (line fluxes of a few $\times
10^{-20}$ Wm$^{-2}$ ) have been reached routinely in the last years
(Hu \etal\ \cite{Hu98},\cite{Hu99},\cite{Hu02}, Rhoads \etal\ \cite{Rhoads03}, 
Maier \etal\ \cite{Maier02}).  The most
distant known object in the universe has been found in this way (Hu
\etal\ \cite{Hu02}).  But still, the number of
Ly-$\alpha$ bright galaxies at $z > 5$ which have been found in
systematic surveys is very limited, and larger samples are
required to draw any firm conclusions about the epoch of galaxy formation.

For both the interpretation of the results of
present narrow-band searches for Ly-$\alpha$  bright galaxies and the
optimum design of future surveys, it is
essential to estimate the expected abundance of these objects under
reasonable assumptions about the cosmological parameters and the history of
galaxy formation.

Here we present a phenomenological model to predict the expected surface
density of Ly-$\alpha $ bright PGs at high redshifts based on ideas which have
already been sketched in Thommes \& Meisenheimer (\cite{Thommes95})
and Thommes (\cite{Thommes96}).

In principle, the prediction of the abundance of Ly-$\alpha$ bright galaxies at high redshifts
 can be obtained in two ways: One way starts from a primeval
density field in the early universe and follows the collapse of dark matter
haloes by N-body simulations or with the Press Schechter formalism. 
Adding in baryonic matter and a reasonable star
formation scenario in combination with a treatment of dust formation and distribution 
could then predict the abundance of star forming haloes
and their star formation rate which -- under the assumption of an initial
mass function (IMF) -- could be converted into a prediction of the number
density of Ly-$\alpha$  bright galaxies above a certain detection limit. 
Haiman and Spaans (\cite{Haiman99}) present calculations along this line.
However, it is not trivial to scale such an {\it ab initio} approach to the observed
abundance of galaxies in the local universe.

Therefore, we pursue the second way in which one tries to extrapolate
the local luminosity function of galaxies and their stellar content
back into the past. This way has been pioneered by Meier
(\cite{Meier76}) and further explored by Baron \& White
(\cite{Baron87}) leading to predicted surface densities of Ly-$\alpha$
bright PGs at $z \simeq 5$ between 1.4 PGs$/ \sq\arcmin /(\Delta z
=1)$ (for $q_0 = 0.5$) and 0.05 PGs$/ \sq\arcmin /(\Delta z =1)$ ($q_0
= 0.0$) for a survey limit $S_{lim} \ge 10^{-19}$\,Wm$^{-2}$. Such a
high abundance has been clearly ruled out by the narrow-band surveys
carried out by Thompson et al. (\cite{Thompson95a}) and more recently
by Hu \etal\ (\cite{Hu98}).  This has triggered several new efforts to
search wider fields to deeper limits (\eg\ the Calar Alto Deep Imaging
Survey -- CADIS see Meisenheimer \etal\ \cite{Meisenheimer97}, \cite{Meisenheimer98}, the
Large Angle Lyman Alpha survey -- LALA, see Rhoads \etal\
\cite{Rhoads00}).  
We have identified two main points which Baron \& White
(\cite{Baron87}) did not take into account: (1) The Ly-$\alpha $
bright phase of PGs might be rather short due to rapid dust formation,
and (2) galaxies show a substantial age spread and therefore did not
form or start simultaneously with their first star formation.  As we will see,
both points tend to reduce the expected number densities so that
the apparent contradiction between observations and predictions disappears.

In the present paper we will describe the basic assumptions of our
models and discuss how choices of the cosmological parameters and the
history of galaxy formation and of primeval star formation would influence the observable
number density for a very broad range of search redshifts $ 3 < z <
13$ and detection limits between $10^{-19}$ and $10^{-21}$
\,Wm$^{-2}$. In a forthcoming paper (Meisenheimer, Thommes and Maier
2004, in the following referred to as paper II) we will try to combine
all available survey results to constrain
the free model parameters even further and thus provide a much 
more confined set of predictions
which could be used as bench mark for future surveys.

The present paper is structured in the following way: In section 2 we
briefly describe the narrow-band imaging technique for detecting Ly-$\alpha $ bright galaxies. 
In section 3
we describe the principle assumptions, parameters and functions of our
model. In section 4 we give a very simplified and transparent
version of our model and demonstrate why the earlier predictions by
Baron \& White (\cite{Baron87}) 
were much too optimistic. Section 5 explores the full
range of model parameters in order to identify those which will most
critically affect the predicted number density of Ly-$\alpha$ bright
primeval galaxies. In section 6, we summarize the generic results of
our model. Section 7 discusses how our results are useful in designing
optimum surveys and gives a first account of how well the model agrees
with the results of present surveys.
This issue will be detailed in the subsequent paper II, in which we will
adjust the free model parameters as close as possible to the results
of all available emission line surveys. This will constrain the range of free
parameters even further.

\section{Narrow band imaging technique to search for high-redshift Ly-$\alpha $ galaxies}

Any emission line survey must aim to map the three-dimensional phase space
of objects $(\alpha, ~\delta, ~\lambda_{obs})$, where $\alpha, ~\delta
$ are the positions on the sky and $\lambda_{obs} = (1 + z) \lambda_{el}$ is
the observed wavelength of the emission line, onto the two-dimensional
detector in an optimum way ($\lambda_{el}$= restframe wavelength of the emission line). Standard observational techniques for
emission line surveys are reviewed by Pritchet (\cite{Pritchet94}).

We concentrate on the narrow-band imaging technique (see e.g.  Hippelein \etal\ {\cite{Hippelein03}, 
Maier \etal\ \cite{Maier02}, Meisenheimer \etal\ \cite{Meisenheimer97}}):
Here a rather narrow wavelength range
$\Delta\lambda$ is selected by
using a narrow-band filter or an imaging Fabry-Perot-Interferometer,
while $(\alpha ,~\delta )$ is directly mapped onto the detector
coordinates $(x,y)$. 
This technique provides two main advantages for
the search of Ly-$\alpha$ galaxies at $z \ge 5$: (a) Since the
entire field of view $\Delta \omega$ is mapped onto the detector,
two-dimensional methods can be used for background determination and
source extraction. In general, they perform much better than the
one-dimensional methods used in analyzing long-slit spectra.  (b) One
can place $\delta\lambda_n$ such that it falls into wavelength regimes
of low and smooth night sky emission. This is of high importance when
searching for Ly-$\alpha$ emission at $z > 4.9$ where about 75\% of
the wavelength range is made rather useless by strong OH-lines in the night
sky.

For the following model predictions of the abundance of Ly-$\alpha$
galaxies we assume $\Delta z  = \Delta \lambda
/ \lambda _{Ly\alpha} =0.1 $  ($\lambda _{Ly\alpha}$=121.57 nm) which is typical for the
narrow-band technique. Three redshift intervals are selected such that
the Ly-$\alpha$ line falls into the best night sky windows around
$\lambda \simeq 705, ~815,$ and 920\,nm (that is $z \simeq 4.8,$ 5.7,
and 6.6). In addition, we present predictions for $z = 3.5$ ($\lambda
= 530$\,nm, V band), $z = 9.3$ ($\lambda = 1250$\,nm, J band), and $z
= 12.6$ ($\lambda = 1650$\,nm, H band).

\section{The basic formalism}

We want to calculate the surface number density of Ly-$\alpha$ emitting PGs on the
sky per solid angle $\Delta \omega $ which have detectable
Ly-$\alpha $ fluxes greater than a certain flux limit $S_{lim}$ and
which have redshifts $z_0$ in an interval $[z_0-\Delta z/2, z_0+\Delta z/2]$.
In a narrow band search for Ly-$\alpha$ emitting objects,
$z_0$ is given by the central wavelength $\lambda _0$ of the narrow band filter and
$\Delta z$ by the band width of the filter $\Delta \lambda $
($\Delta z = \Delta \lambda
/ \lambda _{Ly\alpha} $ ; $\lambda _{Ly\alpha}$=121.57 nm).
Here $\Delta \omega $ refers to the area on the sky which is covered
by the entire survey.
$\Delta \omega$, $\Delta z$ and $z_0$ define a certain comoving volume
$\Delta V_c$ of the universe. The intrinsic 
Ly-$\alpha $ flux of PGs at $z_0$ which produce observable fluxes
 of $S_{lim}$ is given by
\begin{equation}
L_{min}(z_0) \, \, = \, \,  4\pi \, S_{lim} \, d_L(z_0)^2\,
\end{equation}
with the luminosity distance $d_L(z_0)$. We have to
calculate the number of PGs in the volume $\Delta V_c$ with
$L_{Ly\alpha} \ge L_{min}(z_0)$ at the epoch $t_0=t(z_0)$. This number 
depends on:
\begin{itemize} 
\item[(i)] The galaxy formation history, which determines the number of 
galaxies in the volume $\Delta V_c$ which are just in the process of forming and therefore might be in their Ly-$\alpha$ bright
phase at the redshift $z_0$.
\item[(ii)] The evolution of the Ly-$\alpha$ luminosity of the
PGs as a function of time and mass of the PG.
\end{itemize}

\begin{figure}
\centering
\includegraphics[width=8.8cm,clip=true]{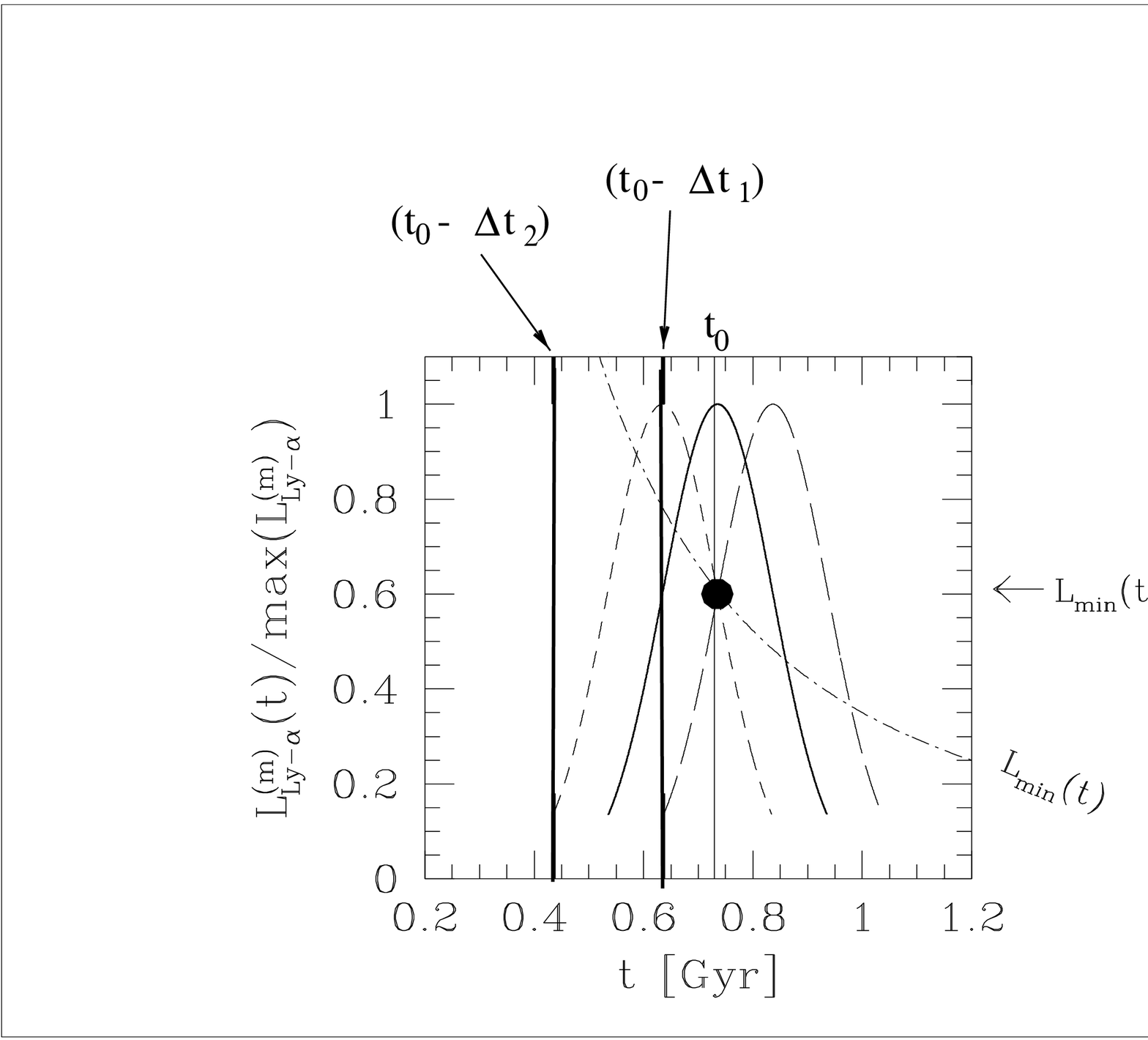} 
\caption[]{ The constraint $t_s$ has to fulfill
so that the Ly-$\alpha $ flux is detectable at
$z_0$. The big black point marks the minimum flux detectable
at $z_0=z(t_0)$ with the detection limit $S_{lim}$.
In order to be detectable in Ly-$\alpha $, the ignition epoch
$t_s$ for the Ly-$\alpha $ emission has to be in the
interval $[t_0-\Delta t_2, t_0 - \Delta t_1]$. This interval is
marked by the thick vertical line.
The short dashed, solid and long dashed curves show the 
possible Ly-$\alpha $ luminosity of PGs as a function
of time with different ignition times $t_s$.
The short dashed curve shows the time evolution of the
Ly-$\alpha $ luminosity of a PG with an $t_s$ at the lower
bound of the relevant time interval. At $t_0$ the Ly-$\alpha $
luminosity of such a PG is already very week e.g. because
of ongoing dust formation. The solid curve corresponds to
a PG with an ignition time $t_s$ centered in the relevant
time interval. It reaches its maximal Ly-$\alpha $ luminosity
at the epoch $t_0$ corresponding to the observed redshift $z_0$.
Therefore, such a PG should be detectable.
On the other hand, the long dashed curve corresponds to 
a PG with a $t_s$ at the right end of the relevant time interval.
The Ly-$\alpha $ luminosity of such a PG has just reached detectable
values and will rise further. }
\label{fig:ptb}
\end{figure}

We assume that the stars in ellipticals and bulges (both
called spheroids) precede the formation of stars in disks which
form out of gas which accretes around the spheroids.
Our hypothesis is that most Ly-$\alpha$ emitting PGs will be these
spheroids during their first burst of star formation at high
redshift (but also refer to our discussion in section 7.4).
We assume that such a ``proto-spheroid'' 
starts to shine in Ly-$\alpha $ as soon
as the first star formation sets in and that the Ly-$\alpha $ 
luminosity is proportional to the star formation rate (SFR) as long
as dust absorption is negligible. Furthermore, it is reasonable
to assume that the SFR of the first star burst is proportional
to the baryonic mass of the PG. Thus, the Ly-$\alpha $ luminosity
of a PG should be proportional to its baryonic mass which we assume to be
proportional to the total mass of the dark matter halo surrounding the
spheroid.
If the SFR is
not constant but changes with time, so too does the Ly-$\alpha $ luminosity.
However, after some time other effects like dust absorption may quench 
the Ly-$\alpha $ luminosity of the PGs, so the proportionality to the
SFR is no longer valid. 
We describe this time dependence of the Ly-$\alpha $ luminosity
with a function $f(t-t_s)$ where $t_s$ is the epoch at which the
galaxy starts to shine in Ly-$\alpha $ (see Fig. \ref{fig:ptb},
we call $t_s$ the ignition time):
\begin{equation}
\begin{array} {c}
L_{Ly_\alpha }^{(M_b)}(t-t_s) = k_{Ly\alpha }~ M_b~ f(t-t_s) \\
\\
  \mbox{with} \quad
  f(t-t_s) ~ \left\{  \begin{array}{cc}
          \equiv  0 & \mbox{if $t < t_s$} \\
                > 0 & \mbox{if $t > t_s $}
                  \end{array}     \right .
\end{array}
\label{eq:Ly-alpha_1}
\end{equation} 
$k_{Ly\alpha}$ gives the factor of proportionality
between SFR and Ly-$\alpha $ flux in a dust-free medium.
$M_b$ denotes the baryonic mass of the PG which we assume to be proportional
to the total mass $M$ of the PG.
$ L_{Ly_\alpha }^{(M_b)}(t-t_s)$ gives the Ly-$\alpha $ luminosity
of a galaxy with mass $M_b$ at the epoch $t$ (time since
the big bang), which started to shine in Ly-$\alpha $ at the
epoch $t_s$.
Because of the assumed proportionality between the Ly-$\alpha $
flux and the PG mass $M$, there exists a lower limit $M_{min}$
for the PGs observed at redshift $z_0$, which can get luminous enough in Ly-$\alpha $
that they have an observable flux $> S_{lim}$.
 PGs with $M<M_{min}$ remain always
too faint in Ly-$\alpha $ to be detectable above the 
survey flux limit $S_{lim}$.

The maximal number of PGs in the volume $\Delta V_c$ which we can
expect to be bright enough in Ly-$\alpha $ is then
given by the integral over the mass function of PGs with lower integration 
limit $M_{min}$ multiplied by $\Delta V_c$:~

\begin{equation}
\int _{M_{min}}^\infty \Phi
(M,z_0) dM~\Delta V_c 
\label{eq:num1}
\end{equation} 

where the mass function $\Phi (M,z_0)dM$ 
gives the comoving
density of objects with mass $M\in [M,M+dM]$ at the observed redshift $z_0$. 

But not all of these galaxies are in the PG phase at the observed epoch $t_0$.
Some of them already left the PG phase or others enter this phase at a later
epoch, so that they are dormant at $t_0$.
The situation is illustrated in Fig. \ref{fig:ptb}.
The Ly-$\alpha$ bright phase during which a PG will be detectable
in Ly-$\alpha$ is constrained to a certain time interval
$[t_s + \Delta t_1(M),t_s+\Delta t_2(M)]$ which is defined by

\begin{equation}
\begin{array}{l}
L_{Ly\alpha }^{(M)}(t-t_s) \, \, \ge \, \, L_{min}(t_0) \\\\
\mbox{for all} 
\, t \in \, [t_s + \Delta t_1(M),t_s+\Delta t_2(M)]
\end{array}
\label{eq:t1t2bestgl}
\end{equation}
The PG is observable in Ly-$\alpha $ at $z_0=z(t_0)$ only if 
$t_0\in [t_s + \Delta t_1(M),t_s+\Delta t_2(M)]$.
This translates to a constraint for $t_s$:
\begin{equation}
t_s \, \,  \in \, \,  [t_0-\Delta t_2(M),t_0-\Delta t_1(M)]
\label{eq:ts}
\end{equation}
Only PGs with $M\ge M_{min}$ which fulfill (\ref{eq:ts}) 
will be detectable in Ly-$\alpha $ at a redshift of $z_0$
(Fig. \ref{fig:ptb}).

The next building block of our model is the 'history of galaxy
formation', that is the distribution of ignition times $t_s$ which we 
assume to be coupled to the formation time of the halo.
Galaxies show a substantial age spread. This translates to
different $t_s$ for different galaxies. Finally our model should end up with
the number of spheroids we observe today, e.g. with the current mass function of
spheroids $\Phi (M,0)$. To take into account the 'history of galaxy formation', we introduce a 
distribution function $P_M(t)dt$, which gives the fraction of current spheroids with mass $M$ which
formed and started their first star formation during the time interval $[t,t+dt]$.
The index $M$ indicates that $P_M(t)$ will in general
depend on the mass of the objects.
We normalize $P_M(t)$ according to
\begin{equation}
   \int_{0}^{t(z=0)} P_M(t) dt = 1
\label{eq:Pmtb}
\end{equation}
$t(z=0)$ is the present age of the universe.\\
$\Phi(M,0) \, P_M(t)dt \, dM $ gives the comoving number density of spheroids with mass in the mass interval
$[M, M+dM]$ collapsing and starting their
first star formation during the time interval $[t, t+dt]$.
The number of PGs per solid angle $\Delta \omega $ with a detectable $Ly-\alpha $ flux above $S_{lim}$
is then given by
\begin{equation}
\begin{array}{l}
N_{Ly_\alpha >L_{min}}(z_0) = \\ \\
\int_{M_{min}(z_0)}^{\infty }
   \left \{ \int _{t_0-\Delta t_2(M)}^{t_0-\Delta t_1(M)} \Phi (M,0) P_M(t) dt \right \}dM
  ~\Delta V_c(z_0)
\end{array}
\label{eq:numlyalpha}
\end{equation}
Some remarks on equation (7):
\begin{itemize}
\item[(i)] In this form, expression (7) is quite general but note that the integration 
limits $t_0-\Delta t_2(M)$ and $t_0-\Delta t_1(M)$ are determined by the
duration of the phase in which the PGs of a certain mass are bright
enough in Ly-$\alpha $ to be detectable. They depend strongly on the
shape of the function $f(t-t_s)$. 
\item[(ii)] The mass function (for spheroids) at redshift $z(t)$ is given by
\begin{equation}
\Phi(M,z(t))= \Phi (M,0) \int _0 ^{t(z)} P_M(t)\, dt.
\end{equation}
With the normalization (\ref{eq:Pmtb}) we ensure that the number of spheroids which form in our model is the number we observe today.
\item[(iii)] Note that with the normalization (\ref{eq:Pmtb}) we assume implicitly that the spheroids which form at high redshift remain until today and therefore we neglect that some spheroids might merge after their formation.
This means that our approach might underestimate the abundance of PGs at intermediate $z$. This could in principle be dealt with by adjusting the right hand side of (\ref{eq:Pmtb}) to values $>1 $. Because the merger rate might depend on the mass of the objects,
these values would clearly be mass dependent. However, in this paper we keep the normalization constraint (\ref{eq:Pmtb})
for simplicity and therefore neglect merging of spheroids after their formation.
\end{itemize}

\section{Discussion of two special cases}

In order to make expression (\ref{eq:numlyalpha}) more transparent, we
choose two special functions for $P_M(t)$ and $L^{(M)}_{Ly\alpha
}(t-t_s)$.  First, we assume for $P_M(t) $ a delta-function $P_M(t) =
\delta (t-t_{in}) $.  That is, all spheroids have the same formation and ignition
epoch $t_s = t_{in}$ and shine in Ly-$\alpha $ simultaneously
(independent of their mass).  If we assume that a fraction $\epsilon $
of the stars of a spheroid are born in the first starburst with a
constant SFR over a period $\Delta t$ (that is $SFR = \epsilon M_b
/\Delta t$, where $M_b$ is the stellar mass $\approx $ baryonic mass of the
spheroid today) and that the Ly-$\alpha $ luminosity is proportional
to the SFR, we get for the Ly-$\alpha $ luminosity
\begin{equation}
L^{(M_b)}_{Ly\alpha }(t-t_s) 
              = \left\{  \begin{array}{cc}
          k_{Ly\alpha } {\epsilon M_b\over { \Delta t}} & \mbox{~if ~$t \in [t_s,t_s +
                                                          \Delta t]$} \\
                 0 & \mbox{otherwise}
                  \end{array}     \right .
\label {eq:lyalphb}
\end{equation}

With this choice for $P_M(t)$ and $L^{(M_b)}_{Ly\alpha }(t-t_s)$
 (\ref{eq:numlyalpha}) gives 
\begin{equation}
\begin{array}{l}
N_{det}(z_0) 
          = \\ \\
\left\{  \begin{array}{cc}
    \int_{M_{min}(z_0)}^{\infty }
    \Phi (M,0)dM
  ~\Delta V_c(z_0) &\, \mbox{ $t_0 \in [t_{in},t_{in}+\Delta t]$} \\
   0 & \,  \mbox{otherwise} \\   
                  \end{array}     \right .
\end{array}
\label{eq:numvisb}
\end{equation}
This is the formula Baron \& White (\cite{Baron87})  used for their calculations.

They took $t_{in} = {1\over 5} t_{coll} $ and $\Delta t = {4\over 5} t_{coll}$,
where $t_{coll} $ is the duration of the collapse associated with a uniform
spherical perturbation of the same initial mean density as the protogalaxy
\footnote{Note that the collapse of a spherical perturbation starts
at the redshift $z_{coll} = z(t_{coll})$.}.  They varied $z_{coll}$
between 6 and 1.5 and took $q_0 = 0.5 $ and $q_0 = 0.05$.  This
corresponds to a variation of $\Delta t $ between 0.6 and 6 Gyrs.  For
the luminosity function $\Phi(L) $ of present-day galaxies they took a
Schechter function with the parameters $\alpha = -1.25$, $L^* =
1.6\times 10^{10}h_0^{-2}$L$_{\odot}$ and $\Phi^* = 1.2 \times
10^{-2}h_0^3$ Mpc$^{-3}$.  Assuming (as we do) that one only can
observe the ancestors of present-day ellipticals and the bulges of
present-day spirals during their violent formation process, they
reduced $\Phi ^*$ by a factor of 3 to $0.4\times 10^{-2} h_0^3$
Mpc$^{-3}$. They used a mass-to-light ratio of 6.6 to convert the
present day luminosity function to a mass function $\Phi(M,z_0)$.  To
convert star formation rates into a Ly-$\alpha $ flux, they used for
the constant $k_{Ly\alpha }$ in (\ref{eq:lyalphb}) the value $k_{BW}
\, :=\, 0.25 \times 10^{35} {\mbox{W}\over {\mbox{m}^2}} {1\over
{\mbox{M}_{\odot} \mbox{yr}^{-1}}}$.  Note that this value is by a
factor 4 smaller than the value deduced from local galaxies with the
assumption of a standard IMF and CASE B recombination
(Kennicutt \cite{Kennicutt83}). In this respect, Baron and White were conservative.
Fig. \ref{fig:verPABW} shows the results we got from
(\ref{eq:numvisb}) for $z_0=4.8$ and $\Delta z =0.1$ with the
parameters of Baron \& White together with the early limits from the
Palomar Fabry-Perot survey for Ly-$\alpha $ emitting PGs (Thompson \etal\ \cite{Thompson95a}, \cite{Thompson95b}) and more recent limits from CADIS (Maier \etal\ \cite{Maier02}) and
LALA (Rhoads \etal\ {\cite{Rhoads03}). Even the least optimistic predictions by
Baron \& White are in obvious conflict with the results of the latest surveys.

\noindent
\begin{figure}
{\includegraphics[width=8.0cm,angle=-90]{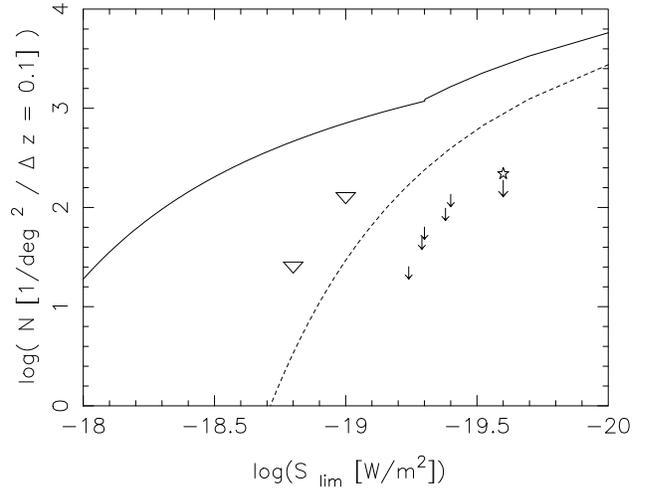} }
\caption[]{{\it Most optimistic (solid line) and pessimistic (dashed line) number 
density predictions for Ly-$\alpha $ emitting PGs
for $z_0=4.8$ and $\Delta z = 0.1$ according to the parameters of
Baron \& White (1987) together with the early upper limits of the Palomar
Fabry-Perot survey of Thompson et al. (\cite{Thompson95a}) (triangles) and more recent searches
by Maier \etal\ (\cite{Maier02}) (small arrows) and Rhoads \etal\ (\cite{Rhoads03}) (open star with arrow).}}
\label{fig:verPABW}
\end{figure}

Alternatively, let us assume that the galaxies 
do not form and start at the same time with their Ly-$\alpha $ bright PG
phase. A simple way to approximate this is to assume that the formation and ignition
times $t_s$ of the galaxies are distributed equally over a certain time
interval. So the delta function for $P_M(t)$ has to be replaced by the
function

\noindent
\begin{equation}
P_M(t)  
 = \left\{  \begin{array}{cc}
   {1\over {t_{out} - t_{in}}} =: {1\over {\Delta t_P}}&
            \mbox{if $t \in [t_{in},t_{out}]$}  \\
                 0 & \mbox{otherwise}
                  \end{array}     \right .
\label {eq:pmc}
\end{equation}  
Furthermore, we assume 
that the genuine Ly-$\alpha $ bright PG phase only lasts
for a limited period, e.g. $\Delta t _{Ly\alpha } = 0.2$\,Gyr 
as discussed above:
\begin{equation}
L_{Ly\alpha }^{(M_b)}(t-t_s) \, =\, \left \{ \begin{array}{ll}
 k_{Ly\alpha } {\epsilon \, M_b \over \Delta t }
  \, \, & \mbox{for $ t \in [t_s,t_s+\Delta t_{Ly\alpha}]$} \\
  0  \, \, & \mbox{otherwise}
\end{array}
\right .
\label{eq:Lytbsp2}
\end{equation}
With (\ref{eq:pmc}) and (\ref{eq:Lytbsp2}) we get
\begin{equation}
\begin{array}{l}
N_{Ly\alpha >L_{min}}(z_0)\, =\\ \\
{\delta t\over {\Delta t_P}} ~
  \int _{M_{min}(z_0)}^{\infty} \Phi (M,0) dM ~ \Delta V_c \\ \\
\end{array}
\label{eq:nlybsp2a}
\end{equation}
with
\begin{equation}
\delta t \le \Delta t_{Ly-\alpha }
\end{equation}

\noindent
Thus, in comparison with (\ref{eq:numvisb}) the numbers are at least reduced
by the factor $\Delta t_{Ly-\alpha}/ \Delta t _P $. In Table \ref{tabfac}
we show this factor for different $z_{in}$, $z_{out}$ ($z_{in}$ and
$z_{out}$ are the redshifts, corresponding to $t_{in}$ and $t_{out}$ in
(\ref{eq:pmc})) and a $\Delta t_{Ly\alpha}$ of $2\times 10^8$
yr for different values of the matter density $\Omega _M$ and the
vacuum energy density $\Omega _V$. 
Obviously the factors by which the numbers are reduced can be quite large.
This simple example demonstrates that the predictions e.g. from Baron \& White (\cite{Baron87})
are likely to be much too optimistic and could well over-predict the
abundance of Ly-$\alpha$ bright galaxies by a factor of 10 or so. 
\bigskip
\noindent

\begin{table}
\caption{Factors  $\Delta t_{Ly\alpha}/ {\Delta t _P} $ for different
$z_{in}$ and $z_{out}$ and $\Delta t_{Ly\alpha }= 2 \times 10^8yr$
(for $H_0 = 70 \, \mbox{km/s Mpc$^{-1}$}$).}
\label{tabfac}
\begin{center}
\begin{tabular}{|c|c|c|c|c|} \hline
\hskip 0.3 cm $  z_{in}$ \hskip 0.3 cm &
\hskip 0.3 cm $z_{out}$ \hskip 0.3 cm &
\hskip 0.3 cm $\Omega _M$ \hskip 0.3 cm &
\hskip 0.3 cm $\Omega _V$ \hskip 0.3 cm &
\hskip 0.3 cm $\Delta t_{Ly\alpha }\over \Delta t_P$ \hskip 0.3 cm 
\\ 
\hline
 20       &     5    & 1.0 & 0.0 & 0.37  \\ 
          &          & 0.3 & 0.0 & 0.23 \\
          &          & 0.1 & 0.0 & 0.16 \\
          &          & 0.3 & 0.7 & 0.20 \\
          &          & 0.1 & 0.9 & 0.12  \\  
\hline
 20       &     2    & 1.0 & 0.0 & 0.12  \\ 
          &          & 0.3 & 0.0 & 0.08  \\
          &          & 0.1 & 0.0 & 0.06  \\
          &          & 0.3 & 0.7 & 0.07  \\
          &          & 0.1 & 0.9 & 0.04  \\
\hline
\end{tabular}
\end{center} 
\end{table}

\section{The detailed model}

The above examples use oversimplified assumptions.
A proper model instead should aim for more realistic functions
$L^{(M)}_{Ly\alpha }(t)$, $P_M(t)$ and $\Phi(M,0)$. 
We deliberately follow a phenomenological approach: We try to parameterize the situation with reasonable functions and try to fix the parameters using both the present day galaxy population and some results from high redshift observations. Note the close 
similarity to the ``semianalytic'' models of galaxy evolution (see e.g. Somerville \etal\ \cite{Somverville01}).

\subsection{The 'history' of galaxy formation}

In the previous section we approximated the function $P_M(t)$ by assuming
that the ignition times $t_s$ of the galaxies are either fixed at a
certain time or equally distributed over a time interval.  Here
we will estimate the  function $P_M(t) $ from the paradigm that
galaxies are assumed to arise from peaks in the density field
$\delta(\vec{x},t) = (\rho(\vec{x},t) - \bar{\rho}) / \bar {\rho}$. In
the following we will omit the spatial coordinate $\vec{x}$.

Consider the density field $\delta (M,t_i)$ at an initial time $t_i$
smoothed with a box containing the mass $M$ and let $\sigma_0 (M,
t_i)$ be the rms variation of this smoothed density field.  We define
the dimensionless fluctuation amplitude
\begin{equation}
\nu (M,t_i) := {\delta(M,t_i) \over {\sigma_0(M,t_i)}}
\end{equation}
In the linear regime the fluctuation amplitude grows proportionally to 
the linear growth factor $D(t)$.
The density field at an epoch  $t>t_i$, therefore, is given by
\begin{equation}
\delta(M,t) = \delta (M,t_i) D(t) / D(t_i)
\label{eq:lin_dens_ev}
\end{equation}

The density contrast grows according to this simple relation until
non-linear effects become important and the region ceases to expand, 
turns around and collapses to form a virialized halo. 
According to the linear
relation (\ref{eq:lin_dens_ev}) the density contrast would have reached 
a critical value $\delta_c$ at the time the halos form. $\delta _c$
can be estimated from the evolution of an isolated spherical over-dense 
region (see e.g. Padmanabhan (\cite{Padmanabhan93})). The value is of order one and depends
weakly on $\Omega _M$. The exact value will not be important for our
considerations.
According to this rule, 
peaks in the initial density field (at epoch $t_i$ ) with a peak
hight of $\nu(M,t_i) = \delta_c D(t_i) / (\sigma_0(M,t_i) D(t_0))$
will collapse at the epoch $t_0$. 
We define the function 
\begin{equation}
\nu_i(M,t) := {\delta _c D(t_i) \over {\sigma_0 (M,t_i) D(t)}}
\end{equation}

We are interested in the number density
$dN_{\mbox{form}}(t)$
of peaks which collapse in a certain time interval $[t_0, t_0+dt]$.
It depends on the distribution of the peak hights, $p(\nu )$.
$p(\nu)d\nu$ gives the probability that a peak has a $\nu$-value in the interval
$[\nu, \nu + d\nu]$.
$p(\nu )$ translates into a distribution of collapse times.
$dN_{\mbox{form}}(t)$ is given by
\begin{equation}
dN_{\mbox{form}} \sim p(\nu_i(M,t)) \, \left | {d\nu_i \over {dt}} \right | \, dt
\end{equation}
from which we obtain of the distribution in collapse times 
\begin{equation}  
P_M (t) = p(\nu _i(M,t)) \left | {d\nu_i \over {dt}} \right | 
\end{equation}
The distribution of collapse redshifts is then given by

\begin{equation}
P_M (z) = p(\nu _i(M,t)) \left | {d\nu_i \over {dt}} \right |\left | {dt\over{dz}} \right | 
\end{equation}

If we assume that we can approximate the fluctuation spectrum by a
power-law
with index $n$, 
we can write the function $\nu_i(M,t)$
\begin{equation}
\nu_i(M,t) = \left ( M\over M_*\right ) ^{n+3\over 6} {D(t_*)\over D(t)}
\end{equation}
$M_*$ is the characteristic mass scale, which collapses at the time $t_*$.
We treat $t_*$ for mass $M_*$ as a free parameter of our model.
In this way we incorporate uncertainties in $\delta _c$ and the
absolute normalization of the fluctuation power spectrum.

If the density fluctuation field is Gaussian, we can calculate
the peak distribution $p(\nu )$ with the formalism given by Bardeen et al.
(\cite{Bardeen}, BBKS). $p(\nu )$ depends on the width of the power spectrum under
consideration. This is described by the BBKS parameter $\gamma $ 
($\gamma =1 $ corresponds to power at a single wavelength
only; lower values indicate a larger range of wavelength).
In the special case of a power-law spectrum with a Gaussian
filtering $\gamma$ is given by (BBKS)
\begin{equation}
\gamma ^2 = {n+3 \over {n+5}}
\label{eq:gamma}
\end{equation}
Assuming a cold dark matter (CDM) model with adiabatic initial
density fluctuations with a power spectrum $\sim k$, 
the power spectrum evolves as the universe expands into 
a power spectrum with effective power-law indices ranging from
-3 on small scales up to 1 on large scales (see BBKS). On galactic
scales the index is n=-2, which is the value we assume in the following.
With Gaussian filtering (\ref{eq:gamma}) gives $\gamma = 0.58$. 

\begin{figure}
\includegraphics[angle=270,width=8.8cm,clip=true]{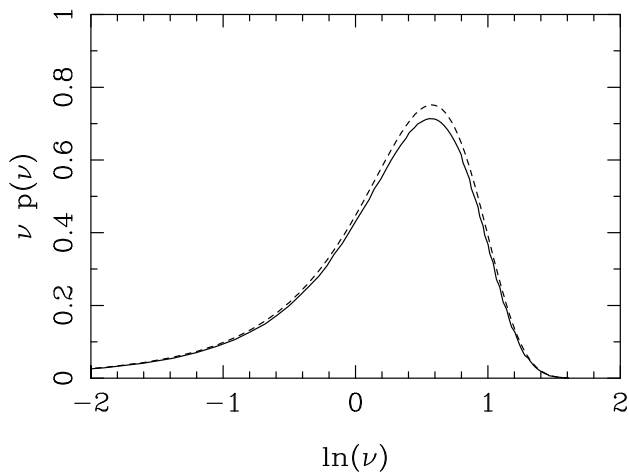} 
\caption[]{{The distribution function $p(\nu)$
for $\gamma = 0.58$ corresponding to an effective power-law index
of the power spectrum of $n=-2$. The solid curve shows the result
of the exact calculation and the dashed one is calculated with the
fitting formula given by BBKS.}}
\label{fig:Pnu}
\end{figure}

Since, in general, $p(\nu)$ cannot be written in closed form,
BBKS give a fitting formula which is accurate enough for our purpose.
Fig. \ref{fig:Pnu} shows $p(\nu)$ for $\gamma = 0.58$
calculated exactly and with the fitting formula provided by BBKS.\\
We normalize $P_M(t)$ according to equation (\ref{eq:Pmtb}).
Fig. \ref{fig:PMt} shows the distribution $P_M(t)$ of the collapse times
for three different masses $0.1*M_*,M_*$ and $5\times M_*$.
Fig. \ref{fig:PMtz} shows the corresponding $P_M(z)$ curves. 
$t_*$ was chosen in such a way that
$P_{M_*}(t(z))$ has its maximum at $z_{\mbox{max}} = z(t_{\mbox{max}}) = 6$.
The curves illustrate again the crucial point of 'bottom up'
hierarchical structure formation: $P_M(t(z))$ with $M<M_*$ peaks
at higher redshifts than $P_{M_*}(t(z))$ and $P_{M}(t(z))$ with
$M>M_*$ peaks at lower redshifts than $P_{M_*}(t(z))$.
Furthermore, the scatter (width) of the collapse times is smaller
for small mass objects which form at high z than for massive objects
which form later. 
Note that the peak formalism is not a strong theoretical motivation for the shape of our function $P_M(t_s)$. By taking $z_{max}$ as a free parameter of the model (see above) we are not strictly applying the peak formalism which would give a definite result for the mass function of halos at a certain redshift once the power spectrum is known. Here we only use the peak formalism to get an estimate for the form of $P_M(t_s)$. Furthermore remember that $t_s$ is not the formation time of the halo but the time when the first star formation starts whereas in the mass function which would be predicted directly by the peak formalism the variable would be the time of collapse $t_{col}$ of the halos. By using $z_{max}$ as a free parameter we leave open how exactly the collapse time and the ignition of star formation in halos are correlated. Comparison of our model with observations may in the end determine wether our choice for $P_M(t_s)$ according to the paradigm of hirarchical structure formation was a good choice or not. 
However, because of the simplicity of our model, we could easily modify the function $P_M(t_s)$ to take into account 
new results such as that bigger galaxies may have earlier star formation epochs as suggested by Heavens et al. (\cite{Heavens04}).
This may be the subject of a following paper.

\begin{figure}
\includegraphics[angle=270,width=8.8cm,clip=true]{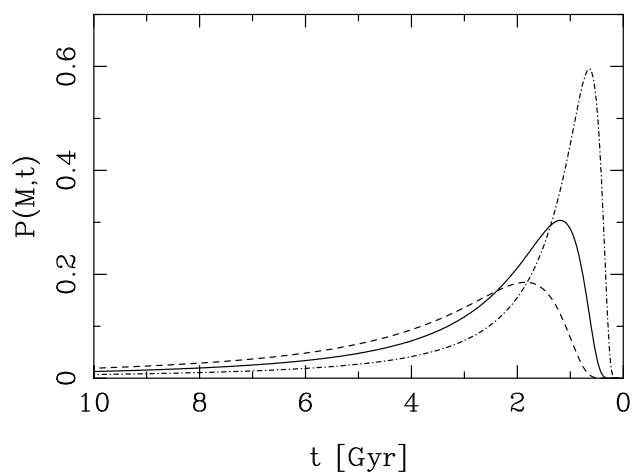} 
\caption[]{{Distribution of collapse times for the three different
masses $0.1\times M_*$ (dotted-dashed line), $M_*$ (solid line) and
$5\times M_*$ (dashed line).}}
\label{fig:PMt}
\end{figure}

\begin{figure}
\includegraphics[angle=270,width=8.8cm,clip=true]{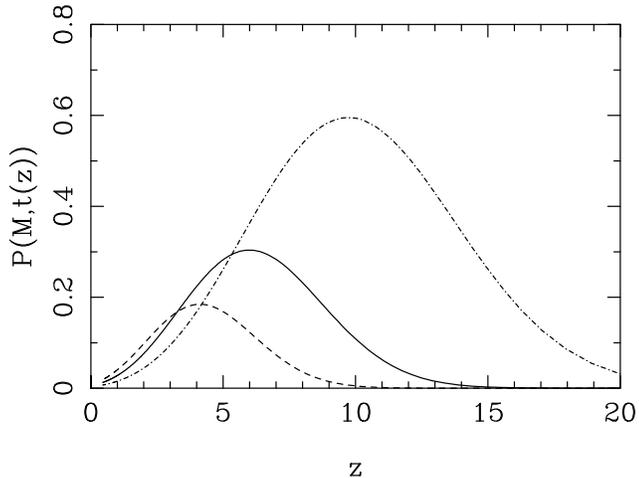} 
\caption[]{{The same curves as in Fig. \ref{fig:PMt}, but with the
time axis transformed into redshift.}}
\label{fig:PMtz}
\end{figure}

\subsection{Development of the Ly-$\alpha$ luminosity with time}

The assumption of a constant star formation rate and Ly-$\alpha $ luminosity
of a PG as in (\ref{eq:Lytbsp2}) is a crude simplification.
Detailed numerical simulations of the formation of galaxies
(\eg\ Steinmetz (\cite{Steinmetz93})) show that the SFR of PGs start with very low values,
increase rapidly and reach a peak after a few $10^8$ years.
The 
metallicity of the early bulge increases rapidly in this phase and
soon reaches
values around 1/10 of the solar value. 
According to our definition this limits the 
genuine Ly-$\alpha$ bright PG phase to a short time period $\Delta t_{Ly\alpha}$
of the order $10^{8-9}$ years. After this time
the metalicity reaches values which inevitably lead to significant dust
formation combined with an absorption of the Ly-$\alpha $ flux.
Although the escape of Ly-$\alpha$ photons from star forming galaxies
is a rather complicated problem depending on several conditions like
the composition of the interstellar medium (e.g. multiphases, 
see Neufeld (\cite{Neufeld91}), or gas flows in the interstellar medium, see Kunth \etal\  (\cite{Kunth99})
, we come back to this points in section 7), 
we simplify and assume that after a certain time period 
the Ly-$\alpha$ emission is no longer proportional to the SFR and decreases mainly due
to the increasing dust content.
We approximate this behavior by a Gaussian for $f(t-t_s)$ 
(see  (\ref{eq:Ly-alpha_1}) )  with a
width $\sigma_{Ly\alpha }$ which remains  a free parameter
of our model:
\begin{equation}
\begin{array}{l}
L^{(M_b)}_{Ly\alpha }(t-t_s) =\\\\
\left\{  
\begin{array}{cc}
   k_{Ly\alpha} \cdot \epsilon \cdot {M_b \over {t_s}} 
  \exp \left \{-{(t-t_s-2\sigma _{Ly\alpha })^2 \over
 {2\sigma_{Ly\alpha }^2}  } \right \} & \mbox{ for $t\geq t_s $} \\
    0                          & \mbox{otherwise}\\
                    \end{array} \right .
\end{array}
\label{eq:Lytbsp3}
\end{equation}
Later on, we will see that the observed number densities of Ly-$\alpha
$ emitting PGs constrains $\sigma_{Ly\alpha }$.  The typical length of
the Ly-$\alpha$ bright phase could be described by the FWHM $\Delta
t_{Ly\alpha} = 2\sigma _{Ly\alpha}\sqrt{2\ln 2}$ of
(\ref{eq:Lytbsp3}).  As already expressed in (\ref{eq:Ly-alpha_1}), we
assume that the star formation rate and therefore the Ly-$\alpha$ flux
during this first phase of star formation is proportional to the
baryonic mass $M_b$ content of the galaxy.  We further assume that
$M_b$ is proportional to the total mass $M$ of the object according to
$M_b = {\Omega _b \over \Omega _M} M$ ($\Omega _b$ is the density
parameter of the baryonic mass in the universe).  Furthermore, we
assume that the star formation rate and accordingly the Ly-$\alpha$
luminosity scales proportionally to $1/t_s$. This is motivated by the
spherical collapse model (see e.g. Padmanabhan,
\cite{Padmanabhan93}). A perturbation which decouples from the cosmic
expansion at the epoch $t_{\mbox{max}}$ will collapse (to a black
hole) at the epoch $t_s = 2\times t_{\mbox{max}}$. The duration of the
collapse process is therefore $\delta t = 0.5\times t_s$.  Hence,
objects which collapse earlier collapse faster and we assume that the
star formation rate is higher as well. Since we do not attempt to
incorporate the complicated processes leading to primeval star
formation and a certain IMF into our model, we summarize this in
the free parameter $\epsilon $, which describes the overall scaling of
the star formation with respect to $M_b$. As in our simple model in
section 4 (see equation (\ref{eq:lyalphb})) , $\epsilon $ can be
interpreted as the factor of proportionality which determines the
fraction of the stars of the spheroid which are born in the first
starburst, although this is not strictly true as
the Gauss function (\ref{eq:Lytbsp3}) describes the combined effect of
star formation and dust production on the Ly-$\alpha $ emission.  If all stars
are produced in a very short starburst, $\epsilon $ could be well above
1.  In this paper we fixed $\epsilon $ to one. This means that the
Ly-$\alpha $ emission of a galaxy at its maximum corresponds to a SFR
of $SFR = {M_b \over {t_s}}$.

\subsection{Mass function}

We assume that
\begin{itemize}
\item Ly-$\alpha$ emitting PGs are the precursors of todays spheroids
\item the baryonic mass we see today in these spheroids is
proportional to the baryonic or total mass of the corresponding PG at
the time of their first star formation 
\item subsequent merging processes or gas accretion does not destroy
the luminous part of the average galaxy, which forms todays spheroid.
However, merging of the PG into a larger halo (e.g. a cluster) may
strip gas and/or the original dark halo of the galaxy and gas
accretion might lead to the formation of a disk. Additional star bursts might
add further stars to the spheroidal system, but the added baryonic mass
is assumed to be proportional to the baryonic/total mass initially present
at the time of the first star burst.
\end{itemize}

In equation (\ref{eq:Lytbsp3}) we therefore need the todays (baryonic) mass function
of spheroids $\Phi _{\mbox{spheroids}}(M_b,0)$.
We derive the local bulge mass function from 
the type dependent luminosity functions
determined from the CFA redshift survey by Marzke \etal\  (\cite{Marzke94}) together with
values for the type-dependent ratio of the bulge to disk luminosity
from Simien \& De Vaucouleurs (\cite{Simien86}) and a constant (baryonic) mass to light
ratio for spheroids set to  $10 M_{\odot }/L_{\odot}$.

Marzke \etal\ (\cite{Marzke94}) described the luminosity functions
for all Hubble types $T$ by Schechter functions
\begin{equation}
\Phi _T (L) dL \, =\, \Phi ^*_T \left ( {L\over L^*_T}\right )^{\alpha _T}
\exp \left \{ -{L\over L^*_T} \right \} \, d\left ({L\over L^*_T} \right )
\label{eq:LSchechter}
\end{equation}
with the parameters listed in table \ref{tab:CfALF}.

\begin{table}
\caption{{\it Schechter function parameter
for the different Hubble types according to Marzke et.al. (1994)}\label{tab:CfALF} }
\begin{center}
\begin{tabular}{|l|c|c|c|}
\hline
Hubble Typ T        &
$\alpha $ & $ L^* $ &  $ \Phi ^*$  \\
&          & $ [10^9 h_0^{-2} L_{\odot }]$ & $ [10^{-3} h_0^3 Mpc^{-3}]$ \\[.2ex]
\hline 
 & & & \\[-3ex]
 E \quad $T \in [-7,-4]$ &
-0.85 & 7.66 &  1.5 \\ 
 S0   \quad $T \in [-3,0] \quad $ &
-0.94 & 4.88 & 7.6 \\ 
Sa-Sb\quad $T \in [1,4] $  &
-0.58 & 4.79 & 8.7 \\ 
Sc-Sd\quad $T \in [5,7] $  &
-0.96 & 5.2 & 4.4 \\ 
Sm-Im\quad $T >7 $ &
-1.87 & 5.1 & 0.6 \\ 
\hline
\end{tabular}
\end{center}
\end{table}

We write the total luminosity $L_{tot}$ of a galaxy
as the sum of the bulge luminosity $L_B$ and the disk luminosity $L_S$:
\begin{equation}
L_{tot} \, =\, L_B \, + \, L_{S}
\label{eq:Ltot}
\end{equation}
Furthermore, we denote with $\gamma _T$ the ratio of the bulge
luminosity to the disk luminosity:
\begin{equation}
\gamma _T \, :=\, {L_B \over L_S }
\end{equation}
The bulge mass of a galaxy of Hubble type T is given by
\begin{equation}
M_B \, = \, \Upsilon _B {1 \over {1 + \gamma _T^{-1}}} L_{tot}
     \, =: \, \Upsilon ^{eff}_T L_{tot}
 \label{eq:mbulge}
\end{equation}
$\Upsilon _B$ denotes the mass-to-light ratio of the bulge
components of present day galaxies, which is relatively independent
of the galaxy type. We took $\Upsilon _B = 10$ M$_{\odot}$/L$_{\odot }$.
Eq. (\ref{eq:mbulge}) defines $\Upsilon ^{eff}_T$, a type dependent
''effective mass-to-light ratio''. 
Table \ref{mleff} lists the values $\Upsilon ^{eff}_T$
for different Hubble types calculated with
the type dependent $\gamma _T$ values of Simien and DeVaucouleurs
(1986)\nocite{Simien86}.
\begin{table}
\caption{{\it Effective bulge mass-to-light ratio $\Upsilon ^{eff}_T$
for the different Hubble types.\label{tab:Leffektiv}}}
\label{mleff}
\begin{center}
\begin{tabular}{|l|c|} \hline
 \hskip 0.5cm Hubble Typ T \quad & \quad $\Upsilon ^{eff}_{T}$
 \hskip0.5cm \\
 & $[M_\odot / L_\odot]$ \\ 
\hline 
\hskip 0.5cm E    \quad $T \in [-7,-4]$ \quad &    10   \hskip 0.5cm \\ 
\hskip 0.5cm S0   \quad $T \in [-3,0] $ \quad &    5.4 \hskip 0.5cm \\ 
\hskip 0.5cm Sa-Sb\quad $T \in [1,4] $  \quad &     2.8 \hskip 0.5cm \\ 
\hskip 0.5cm Sc-Sd\quad $T \in [5,7] $  \quad &     0.48 \hskip 0.5cm \\ 
\hline
\end{tabular}
\end{center}
\end{table}

 \begin{equation}
M _B^*(T) \, = \, \Upsilon ^{eff}_T \, L^*_T
\label{eq:m_B^*}
\end{equation}
With (\ref{eq:m_B^*}) the luminosity functions 
(\ref{eq:LSchechter}) transform into the type dependent
bulge massfunctions
\begin{equation}
\begin{array}{l}
\Phi _T (M_B) dM_B \, = \\
\\
\quad \quad \Phi ^*_T \left ( {M_B\over M_B^*(T)}\right )^{\alpha _T}
\exp \left \{ -{M_B\over M_B^*(T)} \right \} \, d\left ({M_B\over M_B^*(T)} \right )
\end{array}
\label{eq:m_BSchechter}
\end{equation}
The total (baryonic) spheroid mass function of present day
galaxies is given by the sum over the
mass functions of the different Hubble types
\begin{equation}
\Phi _{tot} (M_B)\, dM_B \, = \, \sum _T \, \Phi _T(M_B)\, dM_B
\label{eq:massftotal}
\end{equation}
We will use (\ref{eq:massftotal}) as an approximation for the mass
function $\Phi (M_b,0) $ in equation (\ref{eq:numlyalpha}).

\subsection{Summary of the model}

We constructed a phenomenological model to calculate the expected surface density of Ly-$\alpha$ emitting, high
redshift young galaxies. According to (\ref{eq:numlyalpha}) we have to take into account
the evolution of the Ly-$\alpha$ luminosity as a function of time and mass of 
the galaxy and the galaxy formation history.
We estimated these functions with the following
assumptions:
\begin{itemize}
\item We assume that Ly-$\alpha$ emitting 
PGs are the precursors of the present day
spheroids (=bulges \& ellipticals) in the phase of their first star burst.
\item The first star formation starts in a dust-free environment.
The enrichment of the IGM with dust in the first hundred million
years of star formation leads to a rapid attenuation of the Ly-$\alpha$ emission. 
The time evolution of the Ly-$\alpha$ emission is described by a 
Gaussian with FWHM $\Delta t_{Ly\alpha}$, which gives the typical
duration of the Ly-$\alpha$ bright phase.
\item The SFR and accordingly the Ly-$\alpha$ emission is assumed to be
proportional to the baryonic mass $M_b$ of the PG.
Furthermore, the SFR is assumed to scale with $1\over t_s$, where
$t_s$ is the collapse time: Comparing objects with the same baryonic
mass,  the ones which form their stars at higher redshifts have
higher star formation rates.
\item We get a first guess for the functional form of the distribution of 'ignition times' $P_M(t_s)$
by the distribution of peak hights $p(\nu )$. However, this is a mere ad-hoc assumption with no strong theoretical motivation.
We fix the parameters $M_*$ and $t_*$ so that $P_M(t_s)$
peaks at a certain redshift $z_{\mbox{max}}$ 
for $M = M_*=4\times 10^{10}$M$_\odot$,
which corresponds roughly to the (baryonic) bulge mass of our
Milky Way. $z_{\mbox{max}}$ is a free parameter of our models.
\item We normalized $P_M(t_s)$ so that our model ends up at $z=0$
with the mass function of spheroids we see today.
This assumes that the spheroids which formed in the early
universe stay as entities until today and that their current baryonic
 mass is proportional to the baryonic mass at the time of their 
first star formation. This does not exclude later merging processes
of the dark matter halos and e.g. accretion of gas, which may lead
to the formation of a disk.
\end{itemize}
An overview of parameters of our model 
is given in Table \ref{tab:parameters}.
Note that we keep those parameters fixed which are determined by
observations of the local universe or theoretical arguments. The free
parameters of our model have to be determined by the
observational statistics of Ly-$\alpha$ bright PGs. This will be
attempted in paper II. Here we will only point to the most obvious
constraints (see section 7.2) and use them to define a ``basic model''.
In Table 4 the parameter values of this ``basic model'' are summarized.

\begin{table*}
\caption{{\it Parameters of the model and their value range explored in this paper. 
The values of the ``basic model `` are chosen in such a way that the model 
is in agreement with recent observations (see section 7.2 )\label{tab:parameters}}}
\label{modpara}
\begin{center} 
\begin{tabular}{|l|c|l|c|}
\hline
Parameter & Value range & Explanation & ``Basic Model'' \\
\hline
$\Omega _M, \Omega _\lambda $ & $\Omega _M = 1.0$, $\Omega _\Lambda =0.0$ & flat CDM (fCDM)& \\
                              & $\Omega _M = 0.3$, $\Omega _\Lambda =0.0$ & open CDM (oCDM)& \\
                              & $\Omega _M = 0.3$, $\Omega _\Lambda =0.7$ & $\Lambda$ CDM & 
$\Omega _M = 0.3$, $\Omega _\Lambda =0.7$ \\
\hline
h                             &  0.7                                      & Hubble paramter $H_0 = h 100$km /s Mpc$^{-1}$ & 0.7 \\
\hline
 $\Delta t_{Ly\alpha} = 2\sigma _{Ly\alpha}\sqrt{2\ln 2}$   & 0.1 - 2 Gyr & duration of the Ly-$\alpha $ bright phase & 0.35 Gyr \\
\hline
$\epsilon$ & $\epsilon \equiv 1$ & star formation effiency & 1 \\
\hline
$n$                           &  -2.0                 & power law index of the power spectrum &-2.0  \\
\hline
$z_{max}$                     & 3-25                  & redshift, at which the distribution of & 3.4 \\
                              &                       & 'ignition times ' $P_M(t_s)$ peaks for & \\
                              &                       & $M=M_* = 4\times 10 ^{10}\mbox{M}_\odot$  & \\
\hline
$\alpha$,$L^*$, 
$\Phi ^*$, $\Upsilon ^{eff}$  & see Table 2 and Table 3 & parameters describing the luminosity function and& 
see Table 2 and\\
                              &                         & the mass to light ratio for todays spheroids & 
Table 3\\
\hline
\end{tabular}

\end{center}
\end{table*}

\begin{figure*}
 \resizebox{\hsize}{!}{\includegraphics[width=10cm,angle=-90,clip=true]{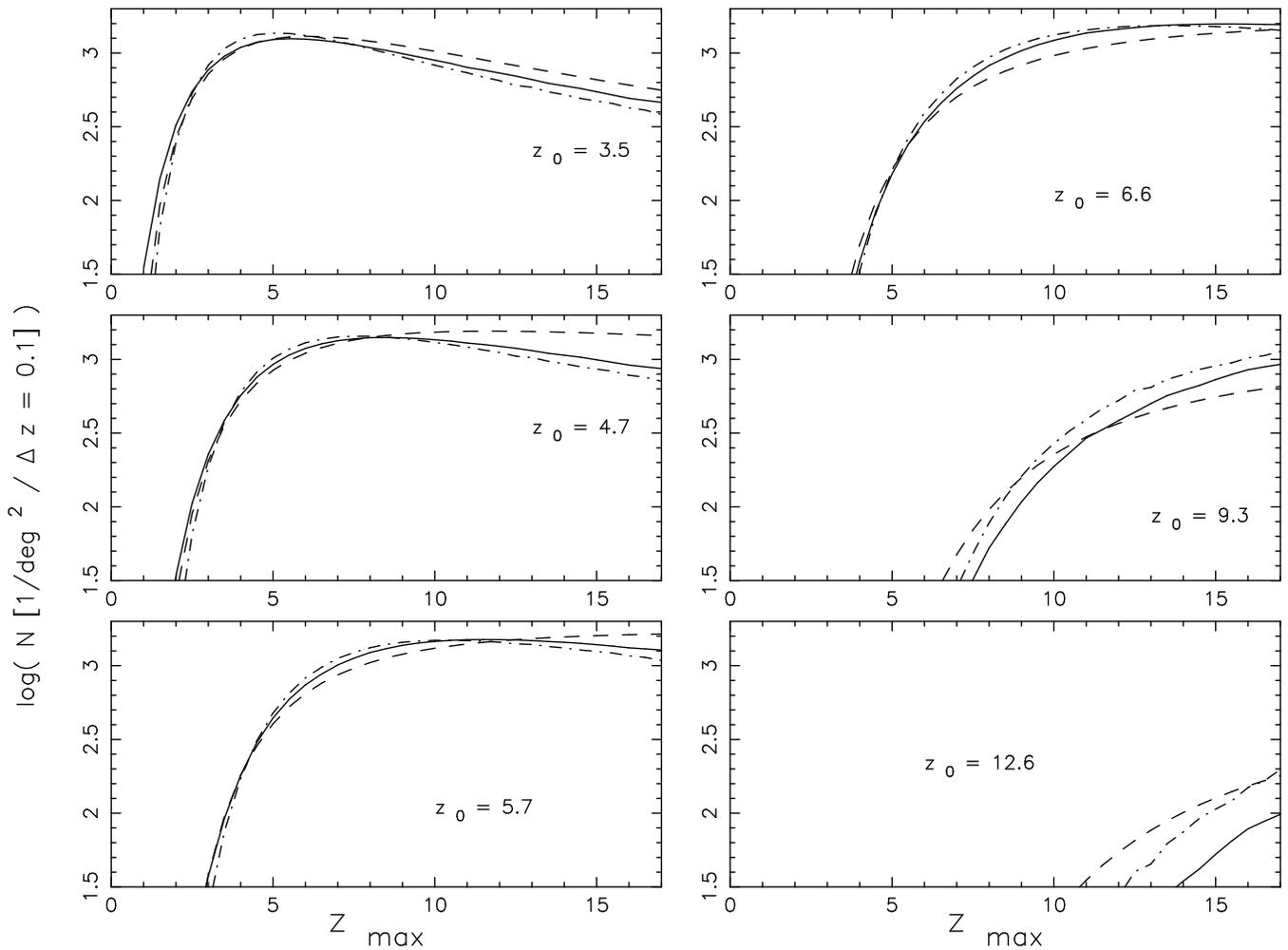} }
 \hfill
\caption{Expected number density per deg$^2$ and $\Delta z = 0.1$ of
 Ly-$\alpha $ emitting PGs with a minimum detectable Ly-${\alpha }$ flux of
 $S_{lim} = 3 \times 10^{-20}$Wm$^{-2}$.  $\Delta t _{Ly\alpha}$ was
 fixed to 0.35 Gyr according to our ``basic model''.  The different panels show the results for
 different observing redshifts $z_0$.  The three different curves in
 each panel correspond to different cosmologies: The solid line
 corresponds to $\Omega _M =0.3$, $\Omega _\Lambda=0.0$, the dashed line
 corresponds to $\Omega _M =1.0$, $\Omega _\Lambda=0.0$ and the dashed
 dotted line corresponds to $\Omega _M =0.3$, $\Omega _\lambda=0.7$.}
\label{fig:logN_zmax}
\end{figure*}

\section{Predicted abundance of Ly-${\alpha }$ galaxies}

Fig. \ref{fig:logN_zmax} shows the expected number density of
Ly-$\alpha $ emitting PGs as a function of $z_{\mbox{max}}$.  The
different panels show the results for different observing redshifts
$z_0$.  Note that the curves for the different redshifts look very
similar.  They rise steeply at small $z_{max} < z_0$ to a maximum and
then fall off slowly at high redshifts $z_{max} > z_0$.  This
behavior can be understood by Fig. \ref{fig:Pzb} which shows $P_M(z)$
for elliptical galaxies with $M=M_*$ and various $z_{max}=3,~6,~9$ and
12.  The two vertical solid lines mark the interval (\ref{eq:ts}) for
$z_0=3.5$.  The number of observable Ly-$\alpha$ emitting PGs is
proportional to the shaded area under the $P_M(z)$ curve in this
interval.  The observed number density is governed by two effects: For
low $z_{max}$ the maximum of the $P_M(z)$ curve lies on the left hand
side of the interval (\ref{eq:ts}), that is at redshifts below $z_0$
(see the solid curve in Fig. \ref{fig:Pzb}).  With decreasing
$z_{max}$ the width of the $P_M(t)$ curve increases rapidly.  This
corresponds to an increasing spread of the ignition times, that is an
increasing $\Delta t_P$ in (\ref{eq:pmc}) (see also the discussion
there).  Taking into account the normalization constraint 
(\ref{eq:Pmtb}), the overall amplitude of $P_M(z)$ decreases
rapidly with decreasing $z_{max}$.  This explains the steep decrease
of the curves in Fig. \ref{fig:logN_zmax} for decreasing $z_{max} <
z_0$.

If $z_{max}$ lies at higher redshift than the interval (\ref{eq:ts})
we face the following situation: With increasing $z_{max}$ the overall
amplitude of $P_M(z)$ increases rapidly (see the dashed, dotted and
dashed dotted lines in Fig. \ref{fig:Pzb}).The reason for this is
that the width of the curves $P_M(t)$ decreases with increasing
$z_{max}$. However, because the total number of galaxies which form is
conserved (see again (\ref{eq:Pmtb})) the total area under the
$P_M(t)$ curve does not change much.  
Thus, the peak amplitudes of $P_M(t)$ and $P_M(z)$
increase. This almost completely compensates for the effect that for
$z_{max} > z_0$ the observed redshift interval (see Fig. \ref{fig:Pzb})
samples lower and lower relative levels of the $P_M(z)$ curve. If one
does not consider one specific mass only but the full range of masses
$M > M_{lim}$, the situation becomes somewhat more involved but the basic
explanation for the behavior of the $N(z_{max})$ curves remains valid.

The lines in the different panels
of Fig. \ref{fig:logN_zmax} reach roughly the same maximum value
$N_{max}$, almost independently of the observation redshift $z_0$.
For instance, for $\Omega_m, \Omega_\Lambda = 0.3,0.7$, we find $1200 <
N_{max} < 1600 /\sq\degr/(\Delta z = 0.1)$ out to redshift $z_0 = 9.3$.
This can be understood in the following way: (1) Because we assumed
that the star formation rate and accordingly the Ly-$\alpha $ flux scales
proportional to $1/t_s$ (see eq. (\ref{eq:Lytbsp3})), galaxies of a
given mass are intrinsically brighter in Ly-$\alpha $ if they form at
higher redshift.  (2) As discussed above, the function $P_M(t)$ is
narrower if it peaks at higher redshifts (see Fig. \ref{fig:PMt} and
Fig. \ref{fig:PMtz}) and thus exhibits a higher peak amplitude.
 
(3) In addition one should note that the comoving volume $\Delta V(z)$
is almost independent of $z$ between $z\approx 3$ and $z\approx 6$.

(1) and (2) lead to an increase of PGs above a fixed mass limit with
increasing $z_0$. On the other hand, increasing $z_0$ means
increasing the luminosity distance $d_L$ of the observed objects, and thus
requires higher masses for PGs which are still observable above the
detection limit $S_{lim}$. Since both $t(z)$ and $D_L(z)$ directly depend
on the geometry of the universe, the fact that $N_{max}$ is almost
independent of $z_0$ is not a coincidence but rather a genuine
property of our model.

After to $z_{max}$, the least known free parameter of our model is the
duration of the Ly-$\alpha$ bright phase, $\Delta t_{Ly\alpha}$.
Actually, it depends on the detailed astrophysical conditions during
the formation of the first generation of massive stars in PGs. Most
notable are the initial mass function (IMF) which determines the rate
of heavy element production, the feedback processes which enrich the
interstellar gas, the cooling of the hot gas phase, the topology of
the star forming regions and many more. Therefore it is essential 
to understand how $\Delta t _{Ly\alpha} = 2 \sqrt{2\ln 2} \cdot \sigma_{Ly\alpha}$
influences the predicted number of Ly-$\alpha$ galaxies in our model.

The panels in Fig. \ref{fig:logN_sigma_Lya} show the expected number
density of Ly-$\alpha $ emitting PGs
as a function of $\sigma _{Ly\alpha}$ for
a survey flux limit $S_{\rm lim} = 3\times 10^{-20}\, \mbox{W/m$^2$}$.  From equation
(\ref{eq:Lytbsp3}) we get
for the length $\Delta t$ of the time interval (\ref{eq:ts}) 
\begin{equation}
\Delta t = \Delta t_2 \, - \, \Delta t_1 = 2\, \sigma _{Ly\alpha}
\sqrt{2 \ln \left ( M \over {M_{min}} \right )}
\label{eq:Delta_t2}
\end{equation}
\begin{figure}
 \resizebox{\hsize}{!}{\includegraphics[width=7cm,angle=-90,clip=true]{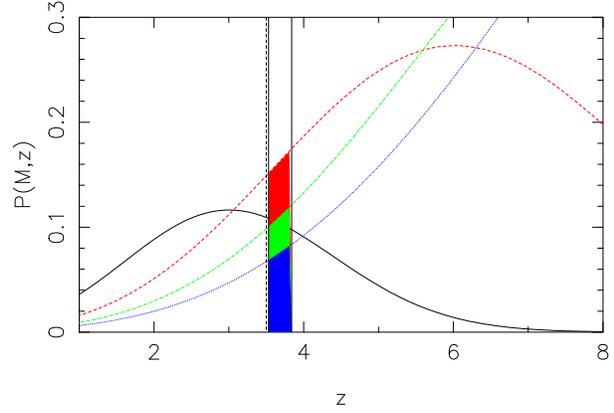} }
 \hfill
\caption{$P_M(t)$ for elliptical galaxies with $M=M_*$  
and different $z_{max}=3,6,9$ and 12.The two vertical lines mark the interval (\ref{eq:ts}) 
for $z_0=3.5$. The dashed vertical line very close to the left border of the interval (\ref{eq:ts})
marks the observing redshfit $z_0=3.5$.}
\label{fig:Pzb}
\end{figure}
First, we consider the case $\sigma _{Ly\alpha} << t(z_0)$, $z_{max}
\ge z_0$. In this case the time interval (\ref{eq:ts}) ends shortly before
$t(z_0)$ and the maximum of the $P_M(t)$ curve is located at higher
redshifts than the redshift corresponding to ($t_0 - \Delta t_2$) (see
Fig. \ref{fig:Pzb}). An increase of $\sigma
_{Ly\alpha }$ has two effects which both increase the expected number
density of Ly-$\alpha $ emitting PGs:
\begin{itemize}
\item The interval (\ref{eq:ts}) over which $P_M(t)$ is integrated increases proportional to 
$\sigma _{Ly\alpha }$
\item The interval (\ref{eq:ts}) moves to the right, that is nearer to the maximum of 
$P_M(t)$.
\end{itemize}  
If $z_{max} < z_0$ and $\sigma _{Ly\alpha} << t(z_0)$ the increase of the interval (5) proportional to 
$\sigma _{Ly\alpha }$ gives an increase of the expected number density, too.
Accordingly, all curves in Fig. \ref{fig:logN_sigma_Lya} 
have a positive slope at small $\sigma _{Ly\alpha }$.
On the other hand, if $\sigma _{Ly\alpha }$ is large, the interval
(\ref{eq:ts}) moves to earlier times before $P_M(t)$ reaches its maximum
and where $P_M(t)$ is a steep function of $t$. This leads to a decrease
of the expected number density of Ly-$\alpha $ emitting PGs when further
increasing $\sigma_{Ly\alpha }$.

\begin{figure*}
\includegraphics[width=11.0cm,angle=-90,clip=true]{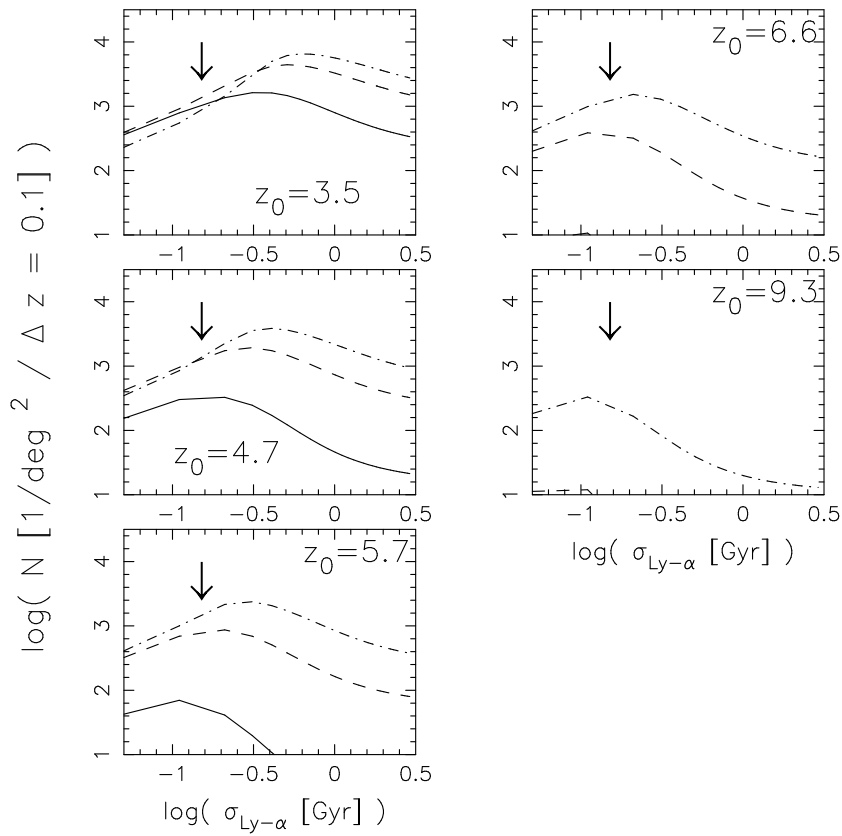} 
 \hfill
\caption{Expected number density per deg$^2$ and $\Delta z = 0.1$ of
Ly-$\alpha $ emitting PGs as a function of $\sigma _{Ly\alpha}$ (see
\ref{eq:Lytbsp3}) for fixed $S_{lim} = 3\times 10^{-20}
\mbox{W/m$^2$}$.  The cosmology is fixed to $\Omega _M =0.3$, $\Omega
_\Lambda=0.7$. The solid curves correspond to $z_{max}=3.4$, the dashed curves
to $z_{max}=6$ and the dotted dashed curves to $z_{max}=10$.The arrow marks the value
$\sigma_{Ly-\alpha}=0.15$ of our ``basic model''.}
\label{fig:logN_sigma_Lya}
\end{figure*}

\begin{figure*}
 \resizebox{\hsize}{!}{\includegraphics[width=5cm,angle=-90,clip=true]{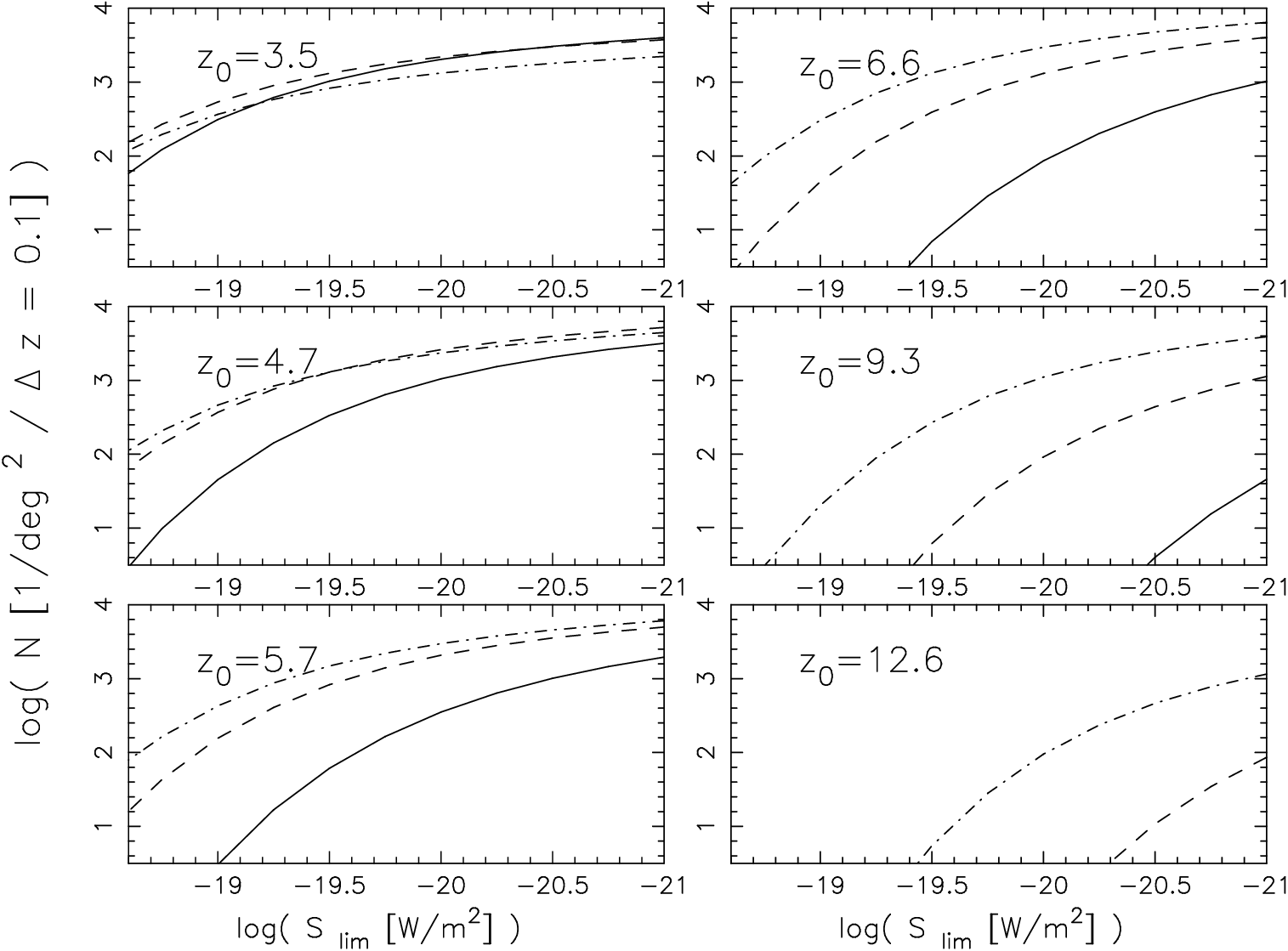} }
 \hfill
\caption{Expected number density per deg$^2$ and $\Delta z = 0.1$ of
Ly-$\alpha $ emitting PGs as a function of the minimum detectable Ly$_{\alpha
}$ flux of $S_{lim}$ according to our ``basic model'' parameters  $\Delta t _{Ly-\alpha}=0.35$ Gyr 
and $\Omega _M =0.3$, $\Omega _V=0.7$. The solid curves
correspond to $z_{max}=3.4$, the dashed curves to $z_{max}=6$ and the
dashed dotted curves to $z_{max}=10$.}
\label{fig:logN_Slim1}
\end{figure*}

Finally, we show the expected abundance of Ly-${\alpha }$ galaxies as
function of the survey limit $S_{lim}$ and observed redshift $z_0$
(see Fig. \ref{fig:logN_Slim1}).  For the purpose of this plot we
fixed the duration of the Ly-${\alpha }$ bright phase to the value of our basic model: $\Delta
t_{Ly\alpha } = 0.35$\,Gyrs (see section 7.2). To
illustrate the dependence on $z_{max}$ we display the cumulative
number density of Ly-$\alpha$ emitting PGs for three different
values of $z_{max}=3.4, 6.0$ and $10.0$. It is obvious from Fig.
\ref{fig:logN_Slim1} that for $z_{max} \la 10$ a set of {\it optical}
surveys for Lyman-$\alpha$ ($z_0 = 3.5$ corresponds to $\lambda({\rm
Ly}\alpha) = 547$\,nm, $z_0 = 6.6$ to 924\,nm) could be sufficient to
determine $z_{max}$. However, when $z_{max}$ is located beyond
$z_{max} = 10$, only the inclusion of a deep near infrared emission
line survey (\eg\ in the J band aiming for Ly-$\alpha$ at $z_0 \simeq
9.3$) would be conclusive. Such a survey seems not feasible from the
ground and would have to wait for the Next
Generation Space Telescope.

\section{Discussion}

After we outlined the generic properties and predictions of our
model, we discuss several aspects which relate 
to current and future surveys for Ly-$\alpha$ emitting primeval
galaxies.

\subsection{Optimum survey strategy}

First, we consider what conclusions about an optimum survey strategy
for Ly-$\alpha$ galaxies can be drawn from the predicted number
densities (Fig.\,\ref{fig:logN_Slim1}):
At bright detection limits (\eg\ $\log(S_{\rm lim}) \ga -19.5$ for $z_0
< 6$ ) the curves
are steep due to the exponential fall-off of the underlying luminosity
function (\ie\ underlying mass function (\ref{eq:massftotal})) towards high
luminosities.  Here it will be more useful to improve the detection
limit of a survey than to enlarge its area. The opposite is true at
very faint detection limits (\eg\ $\log(S_{\rm lim}) \la -20.5$ for $z_0
< 6$). A survey which reaches such a depth will benefit more from an increase
in area than from pushing the limits deeper.

Formally, one might quantify the merit of a Ly-$\alpha $ survey by the
total number $N_{\rm PG}$ of Ly-$\alpha $ emitting PGs which could be found
by spending a given observing time $t_{obs}$ at a given telescope.

Using the observing time $t_{obs}$ to increase the 
observed area $\Delta\omega $ on the sky gives

\begin{equation}
\begin{array}{rl}
N_{\rm PG} \propto \Delta\omega &\propto  t_{obs}\\
&\\
\Rightarrow \log(N_{\rm PG}) &= \log(t_{obs})  + const. \\
&\\
\Rightarrow {\partial \log(N_{\rm PG}) \over \partial \log(t_{obs}) } &= 1 
\label{eq:slope2}
\end{array}
\end{equation}
On the other hand, using observing time to improve the detection 
limit $S_{\rm lim}$:
\begin{equation}
\begin{array}{rl}
S_{\rm lim} &\propto {1\over \sqrt{t_{obs}}}\\
& \\
\Rightarrow \log(S_{\rm lim}) &= -{1\over 2}\log(t_{obs}) \, +\, const.
\end{array}
\end{equation}
If we take into account that $\log(N_{\rm PG})$ is a function 
of $S_{\rm lim}$ 
\begin{equation}
\log(N_{\rm PG}) \, =\, f(\log(S_{\rm lim})) \, =\, f\left (-{1\over
  2} \log(t_{obs}) + const\right )
\label{eq:slope3}
\end{equation}

we find:

\begin{equation}
\begin{array}{rl}
{\partial \log(N_{\rm PG}) \over \partial \log(t_{obs}) } \, &=\, 
f'(\log(S_{\rm lim})) \, {\partial \log (S_{\rm lim}) \over {\partial \log(t_{obs})}}\\
& \\
&= - {1\over 2} f'(\log(S_{\rm lim}))
\end{array}
\label{eq:slope4}
\end{equation}
Notice that $f' < 0$ everywhere (\cf\
Fig.\,\ref{fig:logN_Slim1}). Comparison of (\ref{eq:slope2}) and
(\ref{eq:slope4}) shows how the optimum survey strategy should be
chosen: If $f'(\log(S_{\rm lim})) < -2$ one should use additional observing
time to increase the integration time in order to reach fainter flux
limits.  If $f'(\log(S_{\rm lim})) \ge -2$ additional observing time should
be used to increase the survey area $\Delta \omega $ on the sky in order to
increase the number density of detectable PGs.

\begin{figure}
\includegraphics[width=7cm,angle=-90,clip=true]{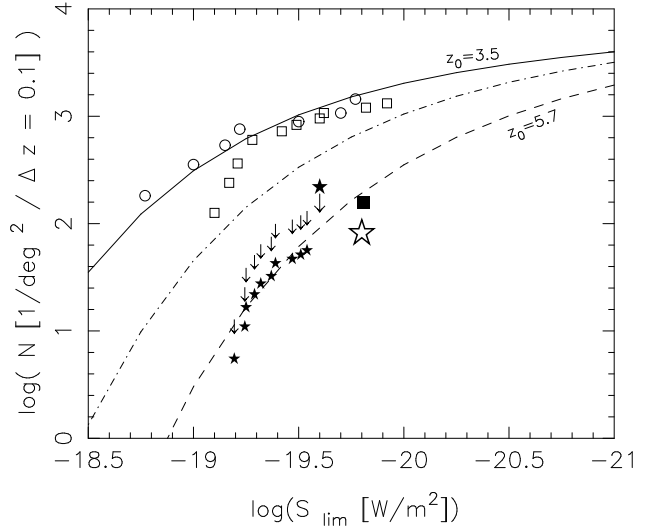} 
 \hfill
\caption{Comparison of our ``basic model'' with observational data.
Big open star: LALA survey of Rhoads \etal\ (\cite{Rhoads03}); filled square: Hu \etal\ (\cite{Hu99}); 
small stars and arrows: CADIS Maier \etal\ (\cite{Maier02});
open boxes: Hu \etal\ (\cite{Hu98});open circles: Kudritzki \etal\ (\cite{Kudritzki00});
filled star with arrow: LALA survey of Rhoads \etal\ \cite{Rhoads03} for $z_0=4.8$;
solid line: $z_0=3.5$; dashed line: $z_0=5.7$; dotted dashed line: $z_0=4.8$;}
\label{fig:best_fit}
\end{figure}

\begin{figure}
\includegraphics[width=7.0cm,angle=-90,clip=true]{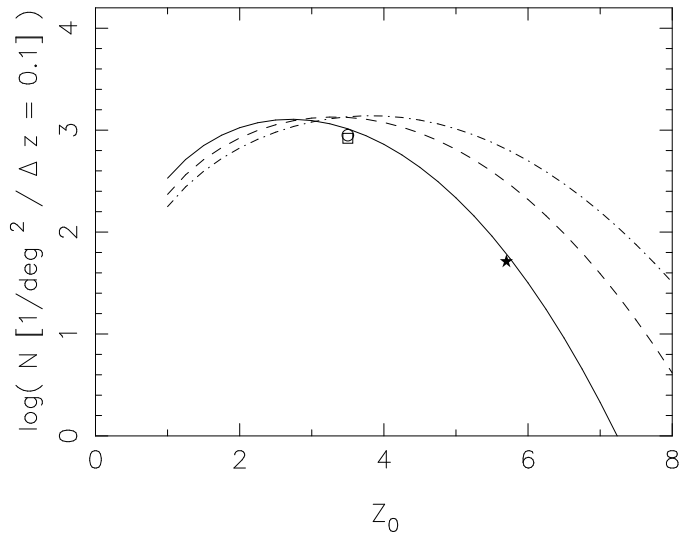}
 \hfill
\caption{Expected number density per deg$^2$ and $\Delta z = 0.1$ of $Ly-\alpha $ emitting PGs
as a function of the observation redshift $z_0$ 
for $\Delta t _{Ly\alpha} = 2\sigma _{Ly-\alpha } \sqrt{2 \ln 2}= 0.35 $ Gyr (according to our ``basic model'')  at a detection limit of $lg(S_{lim} [\mbox{W/m$^2$}]) = -19.5$.
The three curves correspond to $z_{max}=3.4$ (solid curve) ,4.5 (dashed curve), 5.5 (dashed dotted curve). 
The cosmology is fixed to $\Omega _M =0.3$, $\Omega _V=0.7$ (according to our ``basic model'').
The symbols mark observational data: open box: Hu \etal\ (\cite{Hu98}) for $log(S_{lim} [\mbox{W/m$^2$}]) = -19.5$;
open circle: Kudritzki \etal\ (\cite{Kudritzki00}) for $log(S_{lim} [\mbox{W/m$^2$}]) = -19.5$;
small star: CADIS Maier \etal\ (\cite{Maier02}) for $log(S_{lim} [\mbox{W/m$^2$}]) = -19.54$.  }
\label{fig:nz0_zmax}
\end{figure}

\begin{figure}
\includegraphics[width=7.0cm,angle=-90,clip=true]{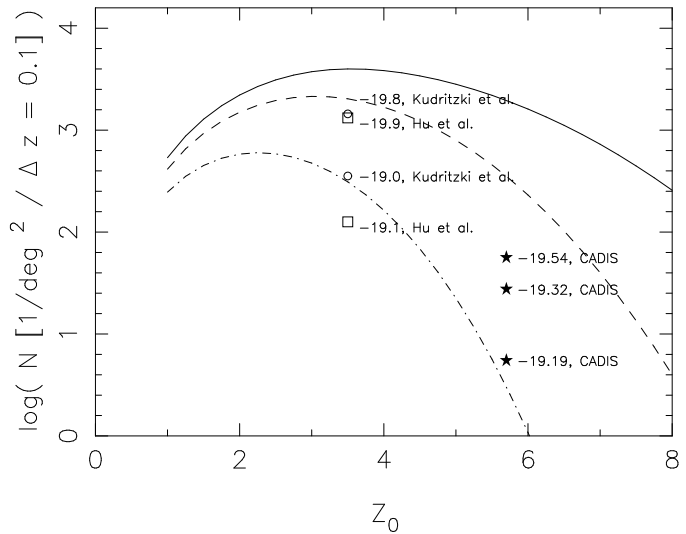}
 \hfill
\caption{Expected number density per deg$^2$ and $\Delta z = 0.1$ of $Ly-\alpha $ emitting PGs
as a function of the observation redshift $z_0$ 
for $\Delta t _{Ly\alpha} = 2\sigma _{Ly-\alpha } \sqrt{2 \ln 2}= 0.35 $ Gyr 
and the detection limits
of $log(S_{lim}[\mbox{W/m$^2$}]) =$ -21.0 (solid line), -20.0 (dashed line), -19.0 (dashed dotted line).  
$z_{max}$ is fixed to 3.4
and the cosmology is fixed to $\Omega _M =0.3$, $\Omega _V=0.7$ (according to our ``basic model''). 
The symbols mark observational data: open box: Hu \etal\ (\cite{Hu98}) for $log(S_{lim} [\mbox{W/m$^2$}]) = -19.1,
-19,9$;
open circle: Kudritzki \etal\ (\cite{Kudritzki00}) for $log(S_{lim} [\mbox{W/m$^2$}]) = -19.0, -19.8$;
small stars: CADIS Maier \etal\ (\cite{Maier02}) for $log(S_{lim} [\mbox{W/m$^2$}]) = -19.54,-19.32,-19.19$.  }
\label{fig:nz0_slim}
\end{figure}

\subsection{Comparison with observed abundance}

We have chosen the parameters of our ``basic model'' in such a way
that our model fits the observed surface density of Ly-$\alpha $
emitting PGs at redshifts $z_0=3.5$ and $z_0=5.7$.
Fig. \ref{fig:best_fit} shows the result.  
If we keep $\epsilon$ fixed to $1$ the main free paramters of the model are 
$z_{max}$ and $\Delta t_{Ly-\alpha }$. These parameters seem to be strongly constrained by the observational data
at different redshifts.
To produce Fig. \ref{fig:best_fit} we varied the two parameters
$z_{max}$ and $\Delta t_{Ly-\alpha }$ by hand in order to fit the data. 
It is non trivial to find a combination of values consistent with all available data.
However, as explained below, part of the difficulty may arise from a second Ly-$\alpha $ bright phase of PGs.\\
Fig.\ref{fig:nz0_zmax}
shows the surface density of Ly-$\alpha $ emitters as a function of
redshifts $z_0$ for different $z_{max}$ and a fixed detection
limit.
Fig. \ref{fig:nz0_slim} shows the same diagram for different
detection limits $\log(S_{lim} [\mbox{W/m$^2$}])=$-21, -20, -19 but
fixed $z_{max}=3.4$. At faint detection limits $\log(S_{lim}
[\mbox{W/m$^2$}]) < -20.0$ the number density peaks at redshifts
$z_0\ge 3$ and falls off slowly with increasing redshift.  For
$z_{max} =$4.5 and 5.5 the surface density of Ly-$\alpha $ emitters at
a detection limit of $\log(S_{lim} [\mbox{W/m$^2$}]) < -20.0$ is
expected to be more or less constant in the redshift range $z_0=3...5$
with values $\ge 10^3$ 1/deg$^2$ / $\Delta z = 0.1$, reaching its
maximum at $z_0 \approx 4$.  For $z_{max}=3.4$ the surface density
falls faster with increasing redshift $z_0 > 3$, especially at lower
detection limits.  However, at faint enough detection limits,
$\log(S_{lim} [\mbox{W/m$^2$}])\approx $-21 (see
Fig. \ref{fig:nz0_slim}), the number density  only changes
moderately in the redshift range between 3 and 5.  At faint detection
limits $\log(S_{lim} [\mbox{W/m$^2$}]) < -20.0$ the expected number
density of Ly-$\alpha $ emitting PGs is expected to be significantly
higher ($\ge \mbox{several}\,\, 100$ / deg$^2$ / $\Delta z=0.1$) out to
very high redshifts of $z_0 \approx 8$.  This may open interesting
prospects for the examination of the epoch of reionisation  
(see Haiman, \cite{Haiman02} and Cen, \cite{Cen03}).
On the other hand according to Fig. \ref{fig:nz0_slim} the surface density of
Ly-$\alpha $ emitting PGs is predicted to be a steep function of
observing redshfit $z_0$ and detection limit at high detection limits
$\log(S_{lim} \mbox{W/m$^2$}])\ge -19$ which explains the failure of
early surveys (see Pritchet \etal\ \cite{Pritchet94}).  

As we determined the model parameters only with the data at
$z_0=3.5$ and $z_0=5.7$ we get an independent prediction for the
surface density of Ly-$\alpha $ emitting PGs at e.g. $z=4.8$ (see
Fig. \ref{fig:best_fit}).}  According to this prediction, the surface
density at $z=4.8$ e.g. at $\log(S_{lim} [\mbox{W/m$^2$}])=-19.5$
should be roughly a factor of 4 higher than at $z=5.7$. But very recent
data by the LALA survey (Rhoads \etal\ \cite{Rhoads03}), CADIS survey
(Maier \etal\ \cite{Maier02}) and the Subaru Deep Field survey
(Shimasaku \etal\ \cite{Shimasaku03}) indicate that the surface
density of Ly-$\alpha $ emitting PGs does not differ much between
$z=4.8$ and $z=5.7$ at $\log(S_{lim} [\mbox{W/m$^2$}])=-19.5$. As we will discuss in
detail in paper II, this can be understood in the framework of our
models if $z_{max}$ is close to 5 or 6 instead of 3.4 (\cf\
Fig. \ref{fig:nz0_zmax} for the case $z_{max}=5.5$). However, 
the low absolute numbers observed at $z_0=5.7$ and $z_0=4.8$ can then 
only be explained by reducing either $\epsilon $ or
$\sigma_{Ly\alpha}$ 
considerably. As this will also reduce the
absolute numbers at $z_0=3.5$ a conflict with the observations is
unavoidable.
This might be a hint
that a large fraction of the Ly-$\alpha $ emitters at $z_0 < 4$ do not
belong to the class of ``primeval'' Ly-$\alpha $ galaxies considered
here, but are galaxies in a later state of evolution (see section 7.4)

\subsection{Comparison with other models}

Besides the calculations of Haiman and Spaans (\cite{Haiman99}) our calculations
are the only ones that try to model the number density of Ly-$\alpha
$ emitters at high redshifts taking into account the formation of dust
during the early phases of star formation and the history of galaxy
formation.  Haiman and Spaans deduced the mass function and formation
history of haloes directly from the power spectrum with the
Press-Schechter formalism, whereas we extrapolated the local luminostiy
function of galaxies and their stellar content back into the past and
only used the power spectrum and peak formalism to deduce a realistic
distribution of formation times. Furthermore, while we use a
phenomenological approach to describe the modulation of the Ly-$\alpha
$ emission by dust formation in the early phase of star formation, Haiman
and Spaans used detailed Monte Carlo simulations of individual
galaxies with a range of masses for the ionizing stars, dust content
and inhomogeneity together with the solutions of the radiative
transfer problem for the Ly-$\alpha $ line in an inhomogeneous
multi-phase medium (Neufeld, \cite{Neufeld91}). 
It is interesting that our model and the model of Haiman and Spaans both
agree in predicting high surface densities of Ly-$\alpha $ emitters out to redshifts $\approx 8$.
Furthermore, both models also predict that the surface density of Ly-$\alpha $ emitters 
as a function of redshift at faint detection limits (see Fig. \ref{fig:nz0_slim})
should be rather flat in the redshift interval between 4 and 6.

\subsection{Lyman-break galaxies and second generation  Ly-$\alpha$ emission.}

In our models we assumed that strong Ly-$\alpha$ emitting PGs are
young spheroids during their very first phase of star formation in
which dust does not play a crucial role. Subsequently the interstellar medium will
be enriched with metals very soon after the onset of star formation.
Due to the ongoing dust formation, Ly-$\alpha$ photons will be more
and more absorbed. The Ly-$\alpha$ flux of the PGs will decrease
although the SFR may still increase and reach its maximum at a
later time.  A Ly-$\alpha$ dark phase follows, during which all
Ly-$\alpha$ photons are destroyed due to resonant scattering in the
dusty interstellar medium and during which the SFR reaches its
maximum.  

At later times after the SFR reaches its maximum, model calculations (see e.g.
Fria\c{c}a and Terlevich (\cite{Friaca99})) predict a phase in which 
strong outflowing winds build up. Once these winds have developed, a second
Ly-$\alpha$ bright phase might develop. 
Complicated outflows of gas with high velocities $v$ are indeed a
common feature of LBGs (see Pettini \etal\ \cite{Pettini98}) and have
also been found in nearby HII galaxies (Kunth \etal\ \cite{Kunth98}).
Large scale outflows not only explain that whenever LBGs show
Ly-$\alpha$ in emission, this is shifted by up to $\approx 1000$km/s
relative to the metal absorption lines but in addition explain their
P-Cygni line profiles.\\
The Ly-$\alpha$
line of objects in this second Ly-$\alpha$ bright phase should be
shifted by the velocity of the outflowing wind relative to the metal
absorption lines and should show a P-cygni profile.  In principle high
resolution high S/N spectra of the objects may allow us to disentangle
Ly-$\alpha$ emitting objects which are in their first or second
Ly-$\alpha$ bright phase.  In our predictions for the number density
of Ly-$\alpha$ emitters we took into account only the first
Ly-$\alpha$ bright phase and neglected the possible second
phase. At very faint detection limits and redshift $z_0 <
4$, Ly-$\alpha$ emitters from the second phase might contribute.
However, because we fixed our parameters to the observed surface
density of Ly-$\alpha$ emitters by Hu \etal (\cite{Hu98}) and the new
CADIS (Maier \etal\ \cite{Maier02}) results, which do not
distinguish between two Ly-$\alpha$ emitting phases of PGs,
our value for $\Delta t_{Ly\alpha}$ might be representative of 
a (weighted) sum of the durations of both phases.  In
this, the duration of the second phase might have a lower weight,
because this phase might be much fainter in Ly-$\alpha$ than the first
one. Furthermore, one would expect that the relative 
``contamination'' by Ly-$\alpha $ emitters in this second Ly-$\alpha $ bright phase increases with
time. This might explain why Ly-$\alpha $ emitters at $z_0=3.5$ seem
overabundant in comparison
to a model which fits their density at redshifts $z_0 > 4.5$.

\section{Conclusions}

We presented a simple phenomenological quantitative model for the expected surface
number density of high redshift Ly-$\alpha$ emitting galaxies.  
We assumed that elliptical galaxies and bulges of spiral galaxies
(which we call spheroids) formed early in the universe while disks
were built up at a later stage.  Thus, we identified the high redshift
Ly-$\alpha$ emitting PGs with these spheroids during their first burst
of star formation. One of the main assumptions of our model is that
the Ly-$\alpha$ bright phase of this first starburst is confined to
the first several hundred million years after the onset of star
formation (duration: $\Delta t _{Ly\alpha}$).  
We assumed an ad-hoc-function for the 
distribution of 'ignition times' with some motivation from the distribution of peak hights in
the peak-formalism.  In order to derive
{\it absolute} number densities, we follow the method backwards in time
as pioneered by Baron \& White (1987):
The number of PGs that form in our model are normalized to the 
present (baryonic) mass function of spheroids.
Using the surface density of Ly-$\alpha$ emitters detected by recent
surveys at redshifts 3.5 and 5.7, we find that the Ly-$\alpha$ bright
phase of primeval galaxies is very likely confined to a rather short
period of $\le 0.5 $ Gyr after the onset of star formation.
Our model predicts that the surface density of Ly-$\alpha$ emitters
with Ly-$\alpha$ fluxes $S_{\mbox{lim}}\le 10^{-20}$W/m$^2$ should be
high ($\ge \mbox{several}\, \,  100/\sq\degr/(\Delta z=0.1)$) out to very high redshifts of
$z_0\approx 8$.  
The substantial number of spectroscopically confirmed high redshift
Ly-$\alpha $ emitting objects at redshifts $z_0\ge 5$ 
(see e.g. Santos et al. \cite{Santos04} or Malhotra and Rhoads \cite{Malhotra04} and references therein)
show that systematic searches for these objects are indeed successful.
Now the main task for observers will be to quantify the selection effects and to separate the second generation Ly-$\alpha $ emitters.
Together with our simple phenomenological model the observation of the Ly-$\alpha $ luminosity function at high
redshifts (e.g. $z=5.7, 6.6, 9.3, 12.6$ as discussed above) may give interesting hints concerning
the peak of galaxy formation activity (from $z_{max}$) and the duration of Ly-$\alpha $ bright phases of PGs 
(from $\Delta t_{Ly-\alpha }$). 
As soon as this is accomplished our model will be able to pin down the formation of galaxies in term of the three parameters
$\epsilon , \Delta t _{Ly-\alpha}$ and $z_{max}$. It will be straightforward to test any physical model of galaxy formation with respect to this intuitive parameterisation of the observed abundance of PGs.

\bigskip
\acknowledgements
ET thanks the Deutsche Forschungsgemeinschaft (DFG) for the grant
which allowed his stay at the Royal Observatory Edinburgh during which
part of this work was completed. The CADIS search for
Lyman-$\alpha$ galaxies is supported by the SFB 439 of the DFG.

\end{document}